%% file: angry.tex
\documentclass[twocolumn,apj]{config/oja}

\usepackage{amsmath}
\usepackage{xcolor}
\usepackage{textgreek}
\usepackage[utf8]{inputenc}
\usepackage[english]{babel}

\usepackage{hyperref}
\hypersetup{
    unicode,
    colorlinks=true,
    linkcolor=linkcolor,
    citecolor=linkcolor,
    filecolor=linkcolor,
    urlcolor=linkcolor,
}
\usepackage{color,colortbl}
\definecolor{linkcolor}{rgb}{0.0,0.3,0.5}
\DeclareGraphicsExtensions{.bmp,.png,.jpg,.pdf}
\usepackage[normalem]{ulem}
\usepackage{multirow}
\usepackage{orcidlink}
\usepackage{placeins}
\usepackage{subcaption}
\usepackage{tabularx}
\usepackage{verbatim}
\usepackage{xspace}
\usepackage{graphicx}

\input{config/fde_emlines}

\input{config/macros}

\graphicspath{ {./figs/} }

\newcommand\blfootnote[1]{%
  \begingroup
  \renewcommand\thefootnote{}\footnote{#1}%
  \addtocounter{footnote}{-1}%
  \endgroup
}
\newcommand{\target}{RUBIES-EGS-9809\xspace}
\newcommand{\monster}{A2744-45924\xspace}
\newcommand{\sdss}{SDSS-J151806.1+424445.1\xspace}
\newcommand{\rosetta}{JADES-GN-28074\xspace}
\newcommand{\targetc}{CAPERS-27615\xspace}
\newcommand{\sign}{\ensuremath{\sigma_\mathrm{n}}\xspace}
\newcommand{\BIC}{\text{BIC}\xspace}
\newcommand\sbullet[1][.5]{\mathbin{\vcenter{\hbox{\scalebox{#1}{$\bullet$}}}}}
\newcommand{\mbh}{\ensuremath{M_{\sbullet[0.85]}}\xspace}
\newcommand{\Lbol}{\ensuremath{L_\mathrm{bol}}\xspace}
\newcommand{\kbolX}{\ensuremath{k_\mathrm{bol,X}}\xspace}
\newcommand{\orcidauthor}[3]{\author{\href{http://orcid.org/#1}{#2$^{#3}$}}}

\begin{document}
\title{\vspace{-0.8cm}A dusty quenching-candidate AGN host at $\MakeLowercase{z}=5.7$: massive quiescent galaxies may quench already during the dust-obscured phase\vspace{-1.2cm}}

\orcidauthor{0000-0003-2388-8172}{Francesco D'Eugenio}{1, 2, *}
\orcidauthor{0009-0009-6418-7154}{Matilde Brazzini}{1, 2, 3, 4}
\orcidauthor{0000-0002-4985-3819}{Roberto Maiolino}{1, 2, 5}
\orcidauthor{0000-0001-5487-2830}{Elena Bertola}{6}
\orcidauthor{0000-0002-6719-380X}{Stefano Carniani}{7}
\orcidauthor{0000-0002-1660-9502}{Xihan Ji}{1, 2}
\orcidauthor{0009-0003-7423-8660}{Ignas Juod{\v z}balis}{1, 2}
\orcidauthor{0009-0006-4990-7529}{Yixiao Liu}{8, 9, 1, 2}
\orcidauthor{0000-0002-8435-9402}{Minjung Park}{1, 2}
\orcidauthor{0000-0002-7392-7814}{Eleonora Parlanti}{7}
\orcidauthor{0000-0001-9820-5773}{Robert G. Pascalau}{1, 2}
\orcidauthor{0000-0003-0736-7879}{Gabriele Pezzulli}{10}
\orcidauthor{0000-0001-6010-6809}{Jan Scholtz}{1, 2}
\orcidauthor{0000-0002-8224-4505}{Sandro Tacchella}{1, 2}
\orcidauthor{0000-0003-1734-8356}{Stefano Zibetti}{6}

\affiliation{$^1$ Kavli Institute for Cosmology, University of Cambridge, Madingley Road, Cambridge, CB3 0HA, United Kingdom}
\affiliation{$^2$ Cavendish Laboratory - Astrophysics Group, University of Cambridge, 19 JJ Thomson Avenue, Cambridge, CB3 0HE, United Kingdom}
\affiliation{$^3$ Department of Physics, Astronomy Section, University of Trieste, Via G.B. Tiepolo, 11, I-34143 Trieste, Italy
}
\affiliation{$^4$ INAF -- Osservatorio Astronomico di Trieste, Via G. B. Tiepolo 11, I-34143 Trieste, Italy}
\affiliation{$^5$Department of Physics and Astronomy, University College London, Gower Street, London WC1E 6BT, UK}
\affiliation{$^6$INAF - Osservatorio Astrofisico di Arcetri, largo E. Fermi 5, 50127 Firenze, Italy}
\affiliation{$^7$Scuola Normale Superiore, Piazza dei Cavalieri 7, I-56126 Pisa, Italy}
\affiliation{$^8$Chinese Academy of Sciences South America Center for Astronomy (CASSACA), National Astronomical Observatories (NAOC), 20A Datun Road, Beijing 100012, China}
\affiliation{$^9$School of Astronomy and Space Science, University of Chinese Academy of Sciences, Beijing 101408, China}
\affiliation{$^{10}$Kapteyn Astronomical Institute, University of Groningen, Landleven 12, NL-9747 AD Groningen, the Netherlands}

\thanks{$^*$E-mail: \sendemail{francesco.deugenio@gmail.com}{Questions about the X-ray galaxy.}{Greetings Francesco,\%0A\%0AHow is life? I would like to kindly ask a clarification about this paper, if I may. First of all, with all the amazing data out there, how come you guys are trawling through these low signal-to-noise stuff? But I digress, sorry. Data quality notwithstanding, I was wondering ... \%0A\%0AWarm regards,\%0A}{francesco.deugenio@gmail.com}}

\begin{abstract}
We present a spectro-photometric analysis of RUBIES-EGS-9809, a broad-line AGN-host galaxy at $z=5.7$
with a Balmer break and a dominant point-source component. The source is clearly dust 
reddened ($\Av \gtrsim 3$ mag), shows hot- and cold-dust emission, and is a 
Compton-thick X-ray emitter, with X-ray-to-bolometric ratio consistent with luminous AGN.
The large \Av agrees with a clear \NaI absorption (EW$=17\pm2$~\AA), possibly 
from a neutral-gas outflow (3-\textsigma), while broad \OIII traces an ionized outflow, 
suggesting AGN feedback may be affecting the host. We estimate a stellar mass 
$\log(\mstar/\Msun)=10.7\pm0.3$, $\Lbol \sim 4\times10^{46}~\ergs$, and a virial
black-hole mass $\log(\mbh/\Msun) \sim 8.1$ (subject to large systematics). The Balmer 
break is spatially resolved (0.1 arcsec, 0.6 kpc), consistent with a stellar origin, 
implying the galaxy is already old in stars while still heavily obscured. This shows 
that massive galaxies can host evolved stellar populations during their compact, 
dust-obscured phase, consistent with formation in an earlier dusty starburst. SED 
modelling suggests a declining SFR, but low-resolution spectroscopy alone cannot 
constrain the recent SFR, so we treat the source as a strong quenching candidate rather
than a secure post-starburst system. Because quenching follows the decline in SFR, 
star-formation-driven outflows are disfavoured, leaving AGN feedback as a plausible 
driver. We speculate that such rapid feedback ($\sim$100 Myr timescale) may precipitate
quenching while the post-quenching phase stays dust-enshrouded, concealing the UV-bright
quenched phase that would otherwise be easily detected, if dust free. By the time dust clears, the UV-luminous 
stars may have already dwindled, leaving a classic dust-free quiescent galaxy. This `dusty path' to
quiescence may explain the lack of intermediate-age quenched progenitors (ages 50--100 Myr) linking 
dusty starbursts to massive quiescent systems at $z=3\text{--}7$.

\end{abstract}


\maketitle

\section{Introduction}
\label{sec:intro}

Massive quiescent galaxies (stellar mass $\mstar \gtrsim
10^{10.5\text{--}11}~\Msun$) at redshifts $z=3\text{--}7$ present a
significant challenge to current models of galaxy formation, because
they are unexpectedly numerous, massive and old compared to theoretical
predictions \citep{carnall+2023a,valentino+2023,long+2024,russell+2024,alberts+2024,baker+2025c}.
Problems with our observations appear unlikely: their quiescent nature is
confirmed by spectroscopy \citep{nanayakkara+2024,baker+2025,
nanayakkara+2025}, their large masses are independently confirmed by
stellar dynamics \citep{carnall+2023}, including from spatially resolved
dynamical models \citep{pascalau+2025}, and their old ages are tightly
constrained by prominent Balmer breaks \citep{carnall+2023,
degraaff+2025b,weibel+2025} or even 4000-\AA breaks
\citep{glazebrook+2024,baker+2025}.
Yet, this remarkable population also presents another challenge:
it is almost as if these galaxies emerged fully formed, with no massive
galaxies with intermediate ages ($\sim$50 Myr) that could plausibly
serve as their progenitors.

The star-forming progenitors should have star-formation rates (SFRs) in
excess of 100~\Msun~\peryr \citetext{e.g., \citealp{carnall+2023,
glazebrook+2024,degraaff+2025b}; although the precise value of the SFR
also depends on the parametrization and probability priors adopted for
the star-formation history, SFH; e.g., \citealp{tacchella+2022a,turner+2025}}. With
such high SFRs, these star-forming galaxies should be readily
detectable in the very same surveys that identified the quiescent
population \citep{carnall+2024,nanayakkara+2024,baker+2025}. Even conservatively assuming that all quiescent galaxies formed in extremely rapid starbursts lasting $\lesssim100~$Myr \citep[which, in reality, applies only to a fraction of $z\sim1\text{--}2$ quiescent galaxies;][]{park+2024}, the observed number density of quiescent galaxies is so high that assuming a visibility ratio of 0.1 still
predicts that we should detect at least several such progenitors within these
survey footprints.  The complete absence of such systems therefore suggests that these progenitors are heavily obscured.
Recent work shows that extreme dust obscuration can be in place already at $z>7\text{--}8$,
with reports of very dusty galaxies and even direct dust detections at sub-mm wavelengths
\citep{sun+2025,bakx+2025,zavala+2025}.
These systems demonstrate that rapid metal and dust build-up is feasible within the first
600~Myr, strengthening the plausibility of an early, deeply obscured formation pathway for the
progenitors of massive quiescent galaxies.

The hypothesis of a highly obscured starburst phase is particularly promising,
because most massive, quiescent galaxies are typically metal rich \citep[e.g.,][]{carnall+2024,
turner+2025}. The build-up of metals and dust and the resulting obscuration
could explain how the star-forming progenitors have
been missed by large-area NIRCam surveys. The effect of dust would be amplified by the compact
nature of these galaxies, whose quiescent descendants have half-light radii as
small as $\re = 250$~pc \citep[e.g.,][]{carnall+2023,carnall+2024,degraaff+2025b}. These compact
starbursts should then be observable via cold-dust emission at sub-mm wavelengths.
But while sub-mm galaxies have been proposed as the missing population, they
seem too extended and rare \citep[sizes of $\re \gtrsim 1$~kpc; e.g.,][]{swinbank+2010,hodge+2016,gillman+2023} to explain the high number density of compact, quiescent galaxies. Important caveats to the size mismatch argument are rapid size evolution \citep{meidt+vanderwel2024,vanderwel+meidt2025}, and central dust in high-mass galaxies \citep{maheson+2025}, which could artificially inflate the UV--optical sizes of
star-forming galaxies relative to their quiescent and dust-poor descendants.
Recent progress in the search for suitable progenitors has come
from the identification of compact, NIRCam-dark galaxies in the
sub-mm survey ASPIRE \citep{sun+2025}.
These galaxies are so dust-obscured that they remain undetected by NIRCam,
yet their relatively low SFRs also make them elusive in shallow sub-mm surveys
\citep{sun+2025}.

However, even these compact, NIRCam-dark galaxies do not fully solve the
`missing progenitor' problem. We still lack the population of intermediate-age
galaxies (50--100 Myr) with weak Balmer breaks and intrinsically strong rest-frame UV continuum,
which should represent the transition phase between dusty starbursts and dust-free
quiescent galaxies. This phase should be easily detectable by \jwst because galaxies
with similar ages but much lower stellar masses ($\mstar = 10^{8.5}~\Msun$) have
already been identified and spectroscopically confirmed \citep{looser+2024,
kuruvanthodi+2024,baker+2025b} at $z=7\text{--}8.5$. Given that this transition
is purely an ageing process, its duration is fixed by stellar evolution, and it is not
sufficiently short to explain the observational gap.

An intriguing possibility is that some compact, dusty galaxies may
already host these intermediate-age populations -- instead of being
active starbursts.
\citet{perez-gonzalez+2025} discovered a Balmer break in the compact, dusty sub-mm
galaxy \textit{Hyde} \citep{schreiber+2018}. This galaxy is found at the centre
of a bright and extended, bi-lobed ionized-gas structure, reminiscent of an outflow.
Likewise, \citet{deugenio+2025c} identified a similar quasar-like nebula at $z=5.89$
around \textit{Mahler}, a $10^{10}$-\Msun galaxy also showing a Balmer break and
luminous UV continuum. While these are viable progenitors, their number density
is at least ten times lower than required.
Although rare, dusty, post-starburst galaxies have also been observed
\citep{setton+2024}, while \citet{ji+2025b} found a ring-like structure of cold
and hot dust around the quiescent galaxy GS-9209 at $z=4.7$. These discoveries
suggest a scenario where massive galaxies stop or slow down their star formation while still
retaining high column densities of dust, which reduces the visibility time of the
`intermediate-age' phase. At some point
after star formation has already started to decline, the dust clears, revealing
an already evolved stellar population.

Since in our hypothesis these galaxies must lack any significant star
formation, active galactic nuclei (AGNs) are the only viable mechanism
for rapid gas and dust removal. Such AGN-driven `outbursts' are supported by tentative evidence of AGNs in both \textit{Mahler} \citep{deugenio+2025c} and \textit{Hyde} \citetext{\citealp{perez-gonzalez+2025}; although in \textit{Hyde} \textit{in-situ} AGNs inside the gas clouds cannot be ruled out, \citealp{perna+2025}}.
In GS-9209, there is even direct evidence of an AGN, detected via broad-line region (BLR) emission \citep{carnall+2023}.

In this article, we identify a galaxy at $z=5.7$ that presents all the main characteristics of this outburst phase, caught in
action. We present a compact, dusty galaxy with an AGN, detected
at all wavelengths from X-rays to radio. The galaxy displays a
Balmer break, which is spatially extended, implying an evolved
stellar population. At the same time, a metal-loaded, multi-phase outflow may be actively removing the dust from this
system, precipitating its transition to the quiescent stage.
After introducing the target and data in Section~\ref{s.data}, we illustrate the methods of our analysis (Sections~\ref{s.imaging} and~\ref{s.anspec}) and the main results (Section~\ref{s.results}). In Section~\ref{s.disc} we discuss our findings and their possible implications for the progenitors of massive quiescent galaxies, while Section~\ref{s.conc} presents a concise summary of our findings and our conclusions.

Throughout this work, we use a flat \textLambda CDM cosmology with $H_0 = 67.4$~\kms~Mpc$^{-1}$ and $\Omega_\mathrm{m}=0.315$ \citep{planck+2020}, giving a physical scale of 6.00~kpc~arcsec$^{-1}$ at redshift $z=5.70$ (all physical scales are given as proper quantities).
Stellar masses are total stellar mass formed, assuming a \citet{chabrier2003} initial mass function, integrated between 0.1 and 120~\Msun.
All magnitudes are in the AB system \citep{oke+gunn1983} and all EWs are in the rest frame, with negative EW corresponding to line emission.

\section{Data and Target}\label{s.data}

Our target (hereafter, \target) was identified as an X-ray emitter in the
`Extended Groth Strip' cosmological field \citep[EGS;][]{rhodes+2000,davis+2007}.

The X-ray detection leverages the deepest \textit{Chandra} X-ray observations available in EGS \citep[AEGIS-XD, 800~ks;][]{nandra+2015}.
We find a spatial match (separation $d<0.29$~arcsec) with the source XID~551 in the AEGIS-XD
catalogue. This source has a 0.5--10~keV X-ray flux of $F = (1.14\pm0.25)\fluxcgs[-15]$ \citep{nandra+2015}, corresponding to
an
observed-frame X-ray luminosity of $L_\mathrm{X} = (4.2\pm0.9)\times 10^{44}~\ergs$ in the 0.5--10~keV band. Applying the $k$-correction to the rest-frame 2--10~keV band (Section~\ref{s.xray}) gives $L_\mathrm{2\text{--}10\,keV} \simeq 2.3\times10^{44}~\ergs$ before absorption correction.

\begin{center}
    \begin{figure*}[!t]
        \centering
    	\includegraphics[width=\textwidth]{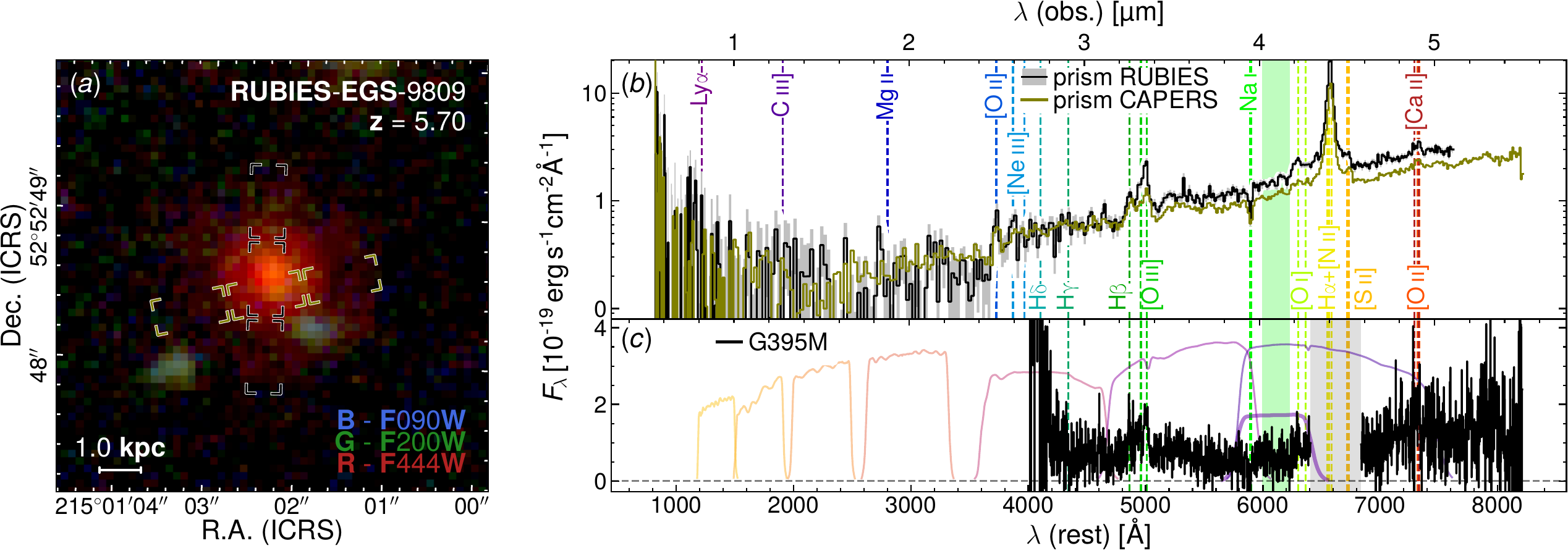}
        {\phantomsubcaption\label{f.data.a}
         \phantomsubcaption\label{f.data.b}
         \phantomsubcaption\label{f.data.c}}
    	\caption{\subref{f.data.a}. NIRCam false-colour image of \target, highlighting the compact nature in
        F444W and the extended emission in the rest-frame UV. The NIRSpec/MSA micro-shutters
        are overlaid for both RUBIES (black) and CAPERS (green). \subref{f.data.b}. Prism spectra, showing
        the Balmer break and broad \Halpha emission,
        while the UV continuum appears very weak, due to strong dust
        attenuation. Weaker rest-optical emission lines are also visible. The optical continuum presents
        both a strong positive slope, indicative of dust reddening, and clear \NaIall absorption; the different
        spectral shape between the RUBIES and CAPERS spectra may be due to the different slit position (panel~\subref{f.data.a}).
        \subref{f.data.c}. \OIIIall emission and \NaIall absorption
        are also visible in the medium-resolution spectrum from the G395M grating, but
        \Halpha falls in the gap between the NIRSpec detectors (grey vertical band).
        Panel~c also shows the transmission curves of the NIRCam filters we use.}\label{f.data}
    \end{figure*}
\end{center}

For the near-infrared (NIR) counterpart, we use \jwst/NIRCam images from the publicly available
survey CEERS \citep[the Cosmic Evolution Early Release Science
Survey;][]{finkelstein+2023}.
The matching NIRCam source is located at $\text{R.A.}=215.01730$,
$\text{Dec.}=52.880158$ and has apparent magnitude 23.3~mag in F444W, but only
27.3~mag in F200W (Fig.~\ref{f.data.a}).
We derive our own NIRCam photometry by fitting the images
(Section~\ref{s.imaging}), and we extend these NIR data to the
mid- and far-infrared (MIR, FIR)
ranges using publicly available data.
Specifically, we use \textit{Spitzer}/IRAC data \citep[as reported by][]{nandra+2015}, \textit{Spitzer}/MIPS 24\mum data from the Far-Infrared Deep Extragalactic Legacy Survey (FIDEL), and a 4-\textsigma detection at 850~\mum by the Submillimetre Common-User Bolometer Array 2 \citep[SCUBA-2;][]{holland+2013,dempsey+2013} on the James Clerk Maxwell Telescope (JCMT), from the SCUBA-2 Cosmology Legacy Survey \citep[source S2CLS~EGS.0113; separation $d=1.5$~arcsec, or 10~percent of the 14.6-arcsec beam size at 850~\mum;][]{zavala+2018}. Finally, we also identify a radio counterpart in the AEGIS20 20-cm survey \citep{ivison+2007}, using the AEGIS20 catalogue \citep{willner+2012}. \target matches EGS20 J142004.19+525248.5, with a separation of 0.3 arcsec (consistent with the stated uncertainty of 0.23 arcsec) and flux density $S_\mathrm{20\,cm}=99\pm13~\mu\mathrm{Jy}$. For the analysis, we use S\'ersic NIRCam photometry (Section~\ref{s.imaging}) and aperture-based \textit{Spitzer}, JCMT, and VLA data; a summary of the non-NIRCam data is provided in Table~\ref{t.phot}.

\input{table_photometry}

In 2024, \target was selected for follow-up spectroscopy by the Cycle 2 programme RUBIES (Red Unknowns: Bright Infrared Extragalactic Survey; PI A. de Graaff and G. Brammer), which targets red sources identified by NIRCam \citep{degraaff+2025a}. The
observations use NIRSpec \citep{jakobsen+2022} and the Micro-Shutter Assembly
\citep[MSA;][]{ferruit+2022}, employing two dispersers: 47 minutes integration with
both the prism and the G395M grating.
In 2025, \target was further observed for 4.74 hours with the prism by the Cycle-3 programme CAPERS
(the CANDELS-Area Prism Epoch of Reionization Survey; PI M. Dickinson; \targetc).
We use data from the DAWN JWST archive \citep[DJA;][]{heintz+2025}, using their v4 data reduction.

The position of the MSA shutters is reported in Fig.~\ref{f.data.a}, with black and green indicating RUBIES and CAPERS, respectively.
The RUBIES observations are better centred on the source, while the CAPERS pointing is offset to the south. For this reason, and to fully benefit from joint
fitting the prism and G395M observations, in the rest of this paper we always use the RUBIES data as reference, except for studying spatial variations in the spectrum.

\section{Imaging analysis}\label{s.imaging}

\subsection{Morphology}\label{s.i.ss.morph}

\begin{center}
    \begin{figure*}[!t]
        \centering
    	\includegraphics[width=\textwidth]{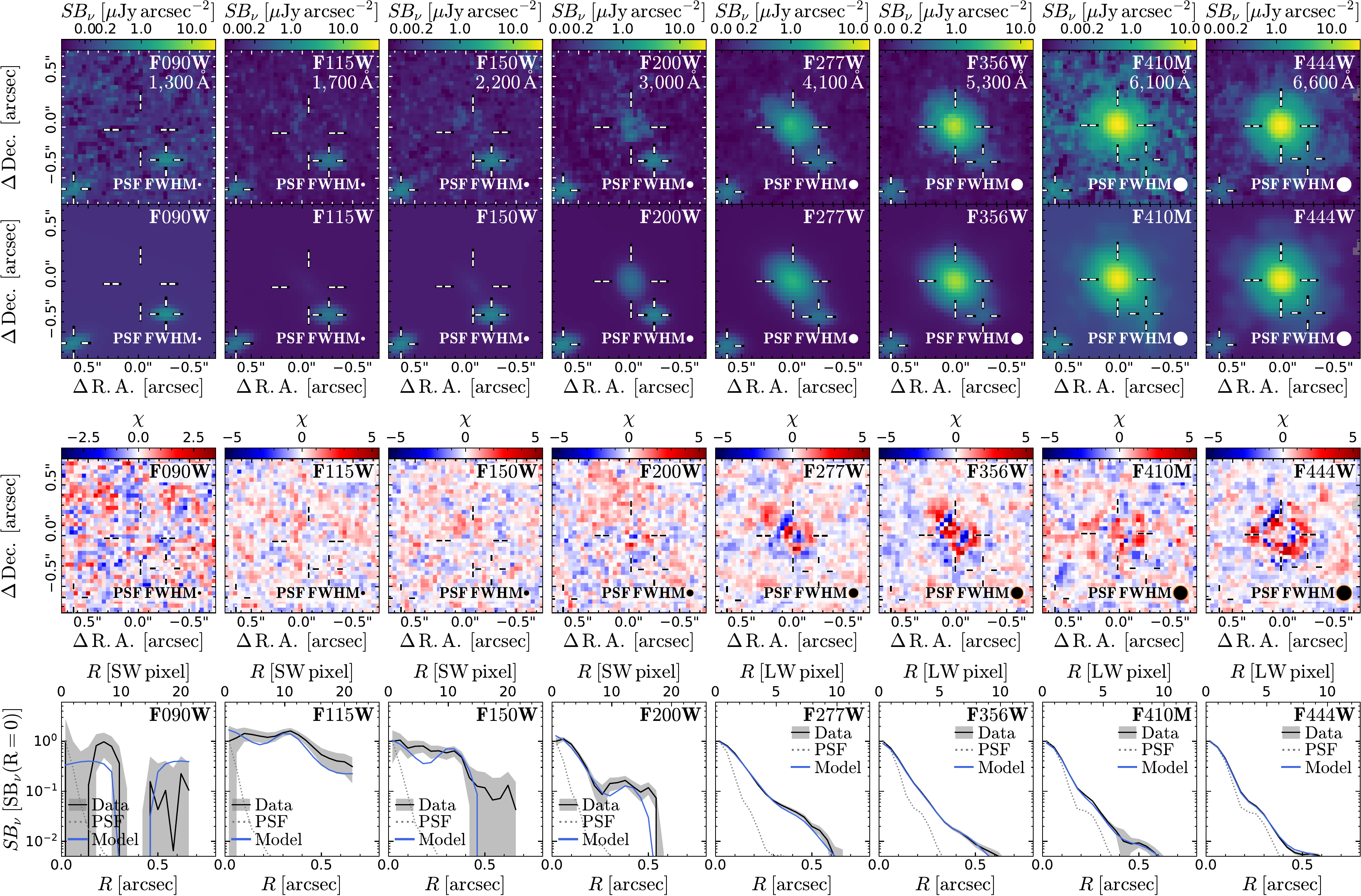}
        \vspace{0.2cm}
    	\caption{NIRCam cutout images (top row), best-fit models (second row),
        $\chi$ residuals (third row; defined by (data$-$model)/noise), and
        radial surface brightness profiles (bottom row). We use an arcsinh scaling
        to enhance faint structures in the cutouts. The NIRCam filter name
        and rest-frame wavelength are indicated in each inset.
        The source morphology transitions from extended in the rest-frame UV
        to marginally resolved in the rest-frame optical.
        }\label{f.sizes}
    \end{figure*}
\end{center}

To measure the morphology and photometry of the source and its immediate environment, we
use the same approach as \citet{deugenio+2025c}, based on multi-component
S{\' e}rsic-profile fitting with \pysersic \citep{pasha+miller2023}.
We obtain 15-arcsec square
cutouts in the eight NIRCam bands observed by CEERS, using the fully reduced mosaics from the Dawn \jwst Archive
\citep[DJA; version v7.2;][]{heintz+2025}. The raw NIRCam images
were processed using a combination of the data-reduction pipeline \textsc{jwst} and \textsc{grizli} \citep{brammer+2023}, as
described in \citet{valentino+2023}. To take into account the
instrument point spread function (PSF), we use empirical PSFs
also provided by the DJA \citep{genin+2025}, and obtained using
\textsc{psfex} \citetext{\citealp{bertin+2013}; following the
methods of \citealp{leauthaud+2007}}. These PSFs are calculated
at the average position angle of CEERS, resulting in a 35\textdegree rotation relative to our cutouts, which we address by de-rotating the PSF accordingly.
The summary of the data and results is shown in
Fig.~\ref{f.sizes} and Table~\ref{t.sizes}. The fiducial model
is the model with maximum posterior probability,
where we used a fat-tailed loss function parametrized by a Student's $t$
distribution. The marginalized posterior probability on the free
model parameters was obtained using the stochastic variational
inference method \citep{hoffman+2013}.

We find three sources within $\sim 2$~arcsec of \target. We mask the brightest one (not shown, just outside the North edge of Fig.~\ref{f.data.a}), and we fit the other two using a S\'ersic profile.
These two sources (tentatively denoted 9809B and 9809C in Table~\ref{t.sizes}) are compact and blue (Fig.~\ref{f.data.a}), but we lack spectroscopic information, so their association with \target remains uncertain.
For \target, our fiducial model is a single S\'ersic profile.

The fit results are reported in Fig.~\ref{f.sizes} and Table~\ref{t.sizes}.
While 9809B and 9809C are detected in all filters, the main source is not
detected in F090W (rest-frame FUV), and only marginally detected in F115W (4-\textsigma
significance). It displays a spatially extended structure in all filters,
more so at blue wavelengths, consistent with a heavily dust-obscured core.
The half-light radius declines steadily from
$\re = 500\text{--}600$~pc in F200W and F277W (rest-frame NUV and B bands, respectively), to only 140~pc in F444W (rest-frame R). The latter is probably
over-estimated, given that the resulting S\'ersic index is $n\sim4$,
suggesting that a point-source component may be necessary in this band.
We also tested a multi-component model consisting of a central point source plus a S\'ersic profile. In this alternative model, all the F444W flux is taken up by the point-source component, although the fit quality does not improve over the simpler single-S\'ersic model, since both models achieve the same reduced $\chi^2$ of 3.5.

\begin{center}
    \begin{figure}[!t]
        \centering
    	\includegraphics[width=\columnwidth]{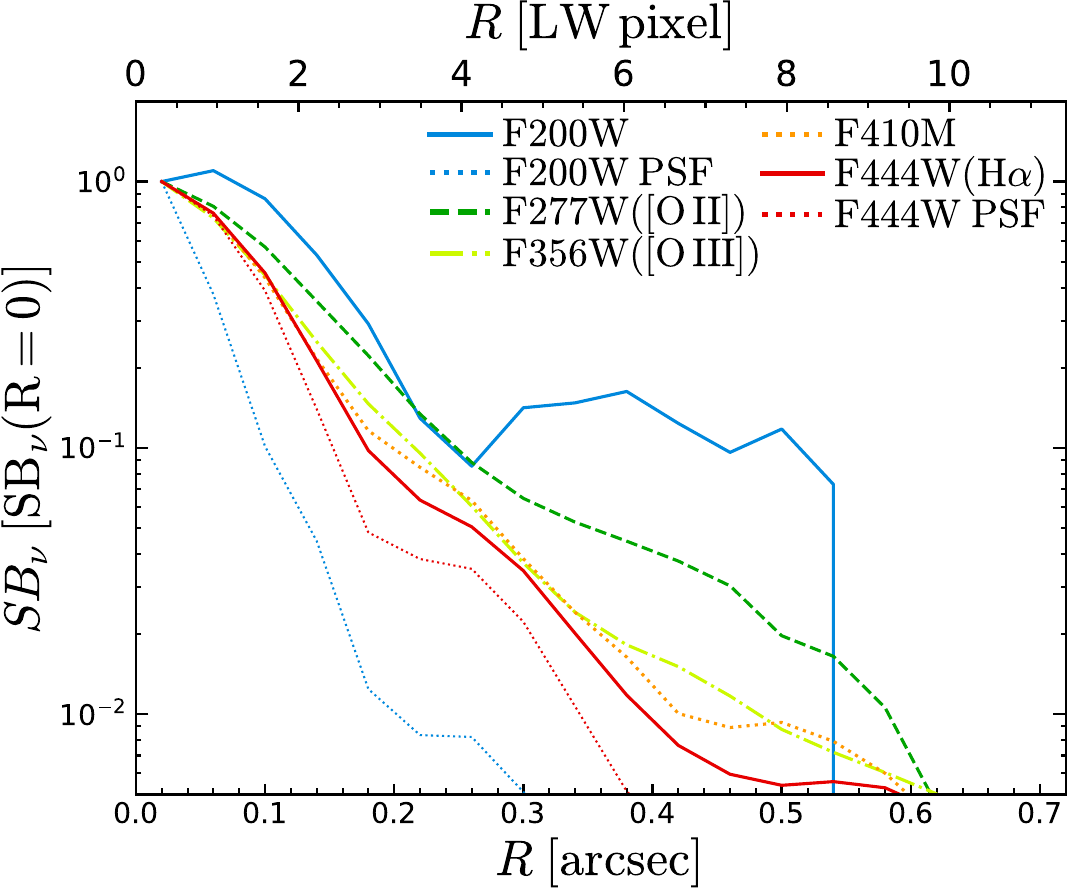}
    	\caption{Observed radial profiles of \target, showing the larger extent at bluer wavelengths (2--2.8~\mum) than at redder wavelengths
        (4.1--4.4~\mum). The F444W PSF is the broadest PSF considered here.
        We also indicate bands containing strong emission lines.}\label{f.sizewave}
    \end{figure}
\end{center}

The fact that a point-source model achieves the same fit quality as the S\'ersic profile may indicate that at 4.4~\mum the AGN is dominating over the host galaxy. This is supported by the bright and broad \Halpha emission line (Fig.~\ref{f.data.b}), and is also consistent with the SED model, where the
AGN starts to dominate redward of $\lambda \simeq 3~\mum$ (Section~\ref{s.i.ss.sed}).
Besides these modelling difficulties, the change in spatial extent with
wavelength is clearly visible in Fig.~\ref{f.sizewave}, where we compare
the observed radial profiles of the five best-detected bands, from F200W to
F444W. The source is clearly more extended at bluer wavelengths,
despite the full-width at half-maximum (FWHM) of the instrument PSF increasing
with wavelength. We also remark that both F277W and F356W contain emission
lines that may be associated with extended gas emission, \OIIall and \OIIIall,
both of which are routinely seen around bright AGNs \citetext{e.g.,
\citealp{crawford+fabian1989,nesvadba+2017,helton+2021,johnson+2022,johnson+2024,
saxena+2024,deugenio+2025a} for \OIIall; \citealp{liu+2013,nesvadba+2017,husemann+2022,perna+2023,
vayner+2023,solimano+2024,saxena+2024,deugenio+2025c} for \OIIIall}.
However, line contamination cannot dominate the F277W morphology. Convolving
the fitted \OIIall and \NeIIIL fluxes from the Balmer-break spectral model
(Table~\ref{t.bbreak}) with the F277W bandpass gives a line contribution of only
$\simeq3$~percent of the observed F277W flux. Even if all of this line emission
were spatially extended, it could not drive the measured F277W half-light radius.

A remaining possibility is that while the continuum is compact, there is diffuse
line emission not captured by the RUBIES aperture spectrum, but still able to
inflate the F277W half-light radius. In such a scenario, the EW of the \OIIall
and \NeIIIL lines increases away from the centre of \target. However, the CAPERS
spectrum, which is offset to the south of \target (Fig.~\ref{f.data.a}), shows
similar or even smaller EWs for both \OIIall and \NeIIIL
(Fig.~\ref{f.data.b}). Thus, while extended ionized gas may contribute at some low
level, the broad-band F277W extension is dominated by continuum emission redward
of the Balmer break.

While emission lines cannot explain the even more extended profile in F200W,
in this filter \target appears double peaked
(Fig.~\ref{f.sizes}; top row, fourth column). Of course,
the double peak (located within $\lesssim 0.1$~arcsec from the peak) cannot be reproduced by our single-S\'ersic model.
At redder wavelengths, the larger PSF and coarser LW sampling may prevent us from separating the two peaks, if they persist. Alternatively, the presence of double-peaked emission at
short observed-frame wavelengths would be in agreement with
the scenario of a heavily dust-obscured core, and is routinely seen in
sub-mm galaxies \citep[e.g.,][]{swinbank+2010,hodge+2016}.

\input{table_sizes_dja_single}

\subsection{SED modelling}\label{s.i.ss.sed}

\begin{figure}
  {\phantomsubcaption\label{f.cigale.a}
   \phantomsubcaption\label{f.cigale.b}
   \phantomsubcaption\label{f.cigale.c}
   \phantomsubcaption\label{f.cigale.d}}
  \includegraphics[width=\columnwidth]{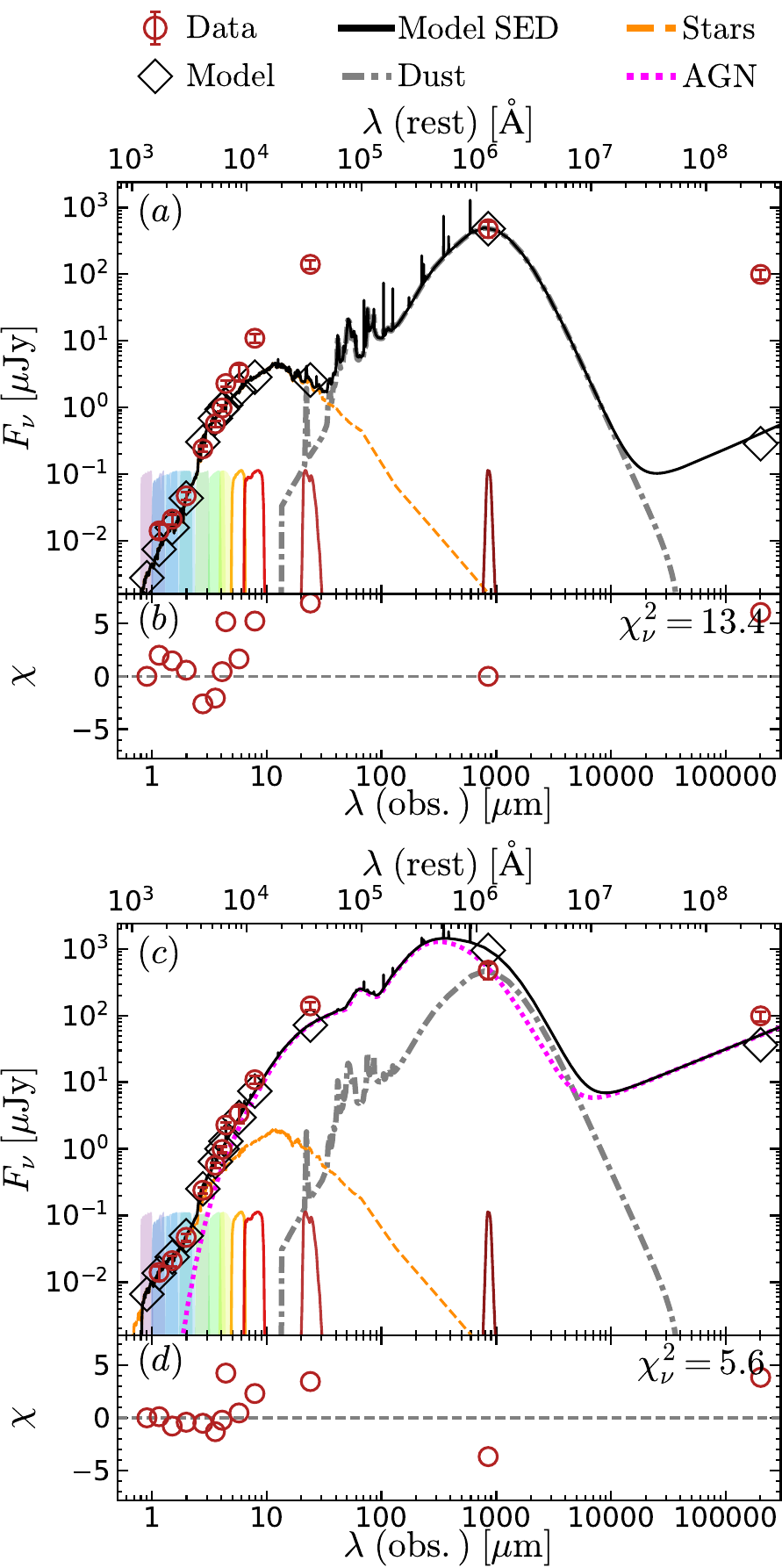}
  \caption{Results of the \cigale modelling, without (panel~\subref{f.cigale.a}) or
  including (panel~\subref{f.cigale.c}) an AGN, now also including the VLA L-band point. The AGN model remains strongly preferred, while both fits imply a declining recent SFH.}\label{f.cigale}
\end{figure}

We use \cigale \citep{yang+2022} to model the galaxy SED, fitting the NIRCam photometry
(from the best-fit models of the
previous section) and the \textit{Spitzer} and JCMT fluxes from the literature.
For the AGN, we employ the \textsc{skirtor} implementation from \citet{stalevski+2016}, while for the host galaxy we use the stellar-population
templates from \citet{bruzual+charlot2003}. We
assume a delayed–exponential star-formation history (SFH), using the standard SFH parametrization implemented in \cigale.
Dust attenuation assumes the \citet{gordon+2003} law for the AGN, and the \citet{calzetti+2000} law for the stellar populations, with extra dust
attenuation towards birth clouds \citep{charlot+fall2000}.
The fit uses the standard \cigale energy-balance framework, so the inferred SFRs
are constrained by the UV--optical stellar light and by the IR emission,
while allowing for a separate AGN contribution to the dust luminosity.
The resulting fits are shown in Fig.~\ref{f.cigale}, where we contrast the
model excluding or including an AGN
(panels~\subref{f.cigale.a}--\subref{f.cigale.b} and panels~\subref{f.cigale.c}--\subref{f.cigale.d}, respectively).

The SED analysis demonstrates the presence of AGN-powered, hot-dust emission, since
models without an AGN cannot match the MIR observations.
Similarly, no amount of star formation can explain the VLA flux.
The model with an AGN is strongly preferred over the model without it, with reduced $\chi^2_\nu=5.64$ and $13.38$, respectively.

From the fiducial model, we derive an AGN bolometric luminosity of
$\log(\Lbol/(\ergs))=46.18$. We do not report the formal \cigale uncertainties for \Lbol, \mstar, $A_V$, or SFR because they are much smaller than, and therefore not representative of, the true error budget, which is likely dominated by systematics such as the adopted SFH parametrization, dust prescription, AGN treatment, and the tension between the fitted photometric points.
We highlight that the fiducial model overpredicts
the FIR flux (4~\textsigma) and underpredicts both the MIR and radio flux (4~\textsigma).
Removing the JCMT/SCUBA-2 and VLA measurements and repeating the fit, we obtain
$\log(\Lbol/(\ergs))=45.9$. We take the difference between this value and the fiducial \Lbol as an estimate of our systematic uncertainties (the formal uncertainties are
only 0.03~dex); within this range,
both results are in reasonable agreement (within 1--2~\textsigma) with what we derive
from emission-line analysis (Section~\ref{s.r.ss.blr}).

We infer total (formed) stellar masses $\log(\mstar/\Msun)=10.72$ (fiducial model)
and $11.24$ (without AGN); the surviving stellar masses are
$\log(\mstar/\Msun)=10.55$ and $11.04$, respectively.
The true uncertainties are dominated by systematics, such as the adopted parametrization for the SFH
\citep[e.g.,][]{leja+2019,carnall+2019}. Recent work has shown that stellar
masses from \cigale can be biased in AGN-dominated systems \citep{buchner+2024}.
In our case, the systematic bias should be $<0.3$~dex \citep[][their
fig.~21]{buchner+2024}. In addition, the detection of a spatially resolved Balmer
break implies that the host galaxy dominates at least at those wavelengths,
providing a robust handle on the underlying stellar populations.

The dust attenuation is consistent between the two models, with $A_V\simeq1.8$~mag
for the continuum and $\Av\simeq4.1$~mag for young stars and emission lines.
As for the SFR, the fiducial AGN model gives $\log(SFR/(\Msun~\peryr))=1.3$ and 1.5
on timescales of 10 and 100~Myr, respectively.
These are still high rates, but they suggest a decreasing trend, which explains the
Balmer break. For reference, the model without AGN has even lower SFRs, with
$\log(SFR/(\Msun~\peryr))=0.9$ and 1.1 on the same timescales.

Before proceeding with the analysis, it is worth recalling that the masses
and SFRs for this kind of source may be particularly challenging to estimate,
which we discuss further in the article (Section~\ref{s.d.ss.notlrd}).

\section{Spectral fitting}\label{s.anspec}

We model the RUBIES prism and grating spectra piecewise, focusing on four spectral features of interest:
the Balmer break, the \Hbeta--\OIIIall emission-line group, the \NaIall doublet,
and the region around \Halpha. The region around the Balmer break is fitted separately,
due to its low signal-to-noise ratio (SNR) compared to the spectral features at
redder wavelengths. When comparing models, we use the Bayesian information
criterion \citep[BIC;][]{schwarz1978}, with a threshold $\Delta \BIC \geq 10$.

The model employs Gaussian lines to model \Hbeta, \OIIIall, \OIall, \Halpha, \NIIall,
\SIIall, and the transauroral quadruplet \OIIAuall. All narrow lines share the same
redshift $z$ and intrinsic velocity dispersion \sign. For \Halpha and \Hbeta, we also
include a broad component, modelled as two Gaussians with a single redshift, but different flux and full-width at half maximum \citep[FWHM;][]{deugenio+2025e}.
For \OIIIall, we also use a broad component with free velocity, representing a gas
outflow (the double Gaussian and the \OIIIall outflow are justified later).
The broad \OIIIall component uses a flat prior on the velocity dispersion $\sigma_\mathrm{out}$
between 0 and 2000~\kms, and is further subject to a complementary-error-function (erfc) prior on $\sign/\sigma_\mathrm{out}$ with mean 1 and scatter 0.1, which avoids swapping the outflow and galaxy components.
The flux ratios of the three doublets with variable line ratio are constrained to their
physically allowed values. Even though we lack the spectral resolution to resolve these
three doublets, our physically motivated priors avoid biasing the model for marginally
resolved doublets \citep[e.g., the \SIIall doublet requires $R\sim450$, which
is reachable by the prism for point sources;][]{degraaff+2024b}.
The other four doublets in the model have fixed line ratios, which we take from
\pyneb \citep{luridiana+2014}. Here we only report the values adopted for the less common
transauroral quadruplet \OIIAuall. The flux of the \OIIL[7320] line is a free
parameter, while the flux ratio between \OIIL[7330] and \OIIL[7320] is free but
constrained within the physically allowed range 0.435--0.565. The remaining two
lines, \OIIL[7319] and \OIIL[7331], arise from the same levels of \OIIL[7330] and
\OIIL[7320], respectively. Therefore, we fix the flux ratios
\OIIL[7330]/\OIIL[7319] = 1.656 and \OIIL[7320]/\OIIL[7331] = 1.908 calculated
again using \pyneb.

\NaIall absorption is modelled using a Gaussian optical depth
with free velocity and velocity dispersion \citep[see e.g.,][]{davies+2024}.
The model employs four free parameters: a variable covering factor $C_f$, the
velocity and velocity dispersion of the absorber, $v_\mathrm{abs}$ and
$\sigma_\mathrm{abs}$, and the optical depth at line centre, $\tau(\NaIL[5896])$.
The optical depth of the \NaIL line is fixed to $2\cdot\tau(\NaIL[5896])$, set
by the oscillator strengths \citep[e.g.,][]{rupke+2005,davies+2024}.

The continuum is modelled piecewise as a polynomial within each of three wavelength
intervals centred on \OIIIL[4959], \NaIL[5896], and \Halpha. The first two intervals
use a 1\textsuperscript{st}-order polynomial, while the third and widest interval
uses a 2\textsuperscript{nd}-order polynomial.
In addition to the background, we add a step function $\mathcal{H}$ under \Halpha,
to account for possible metal absorption (Appendix~\ref{a.metal}).

To reduce the degeneracies in the model, we add external constraints by providing informative
priors, drawn from the sub-mm galaxy ALESS073.1 \citep{parlanti+2024}. We require that the intrinsic narrow-line
dispersion must be $\sign<500~\kms$ \citep[which generously extends the parameter space relative
to the measurement of][i.e., $\sign=220\pm25~\kms$]{parlanti+2024}, and the narrow-line flux ratio $\NIIL/\SIIall$ must be less than 5.
This second prior exploits the sulphur doublet to mitigate the effects
of the otherwise full degeneracy between the narrow \Halpha and \NIIall lines. Of course,
a few cases of higher $\NIIL/\SIIall$ are known \citep[e.g.,][]{deugenio+2024a}, but to
fully test if these conditions apply we would require medium- or high-resolution
spectroscopy.
The outflow velocity dispersion $\sigma_\mathrm{out}$ is penalized against solutions with
$\sigma_\mathrm{out} \leq \sign$; this avoids swapping the narrow and outflow components of
\OIIIall. All three priors are implemented using an erfc function, which penalizes
unwanted solutions more gently than a sharp boundary.

A critical weakness of our model is the inability to estimate an accurate dust attenuation \Av,
due to the lack of \Halpha coverage in the grating. Measuring \Av is crucial,
because the observed spectrum shows clear evidence of dust reddening. We attempt to estimate
\Av from the Balmer decrement observed in the \emph{total} \Halpha and \Hbeta lines, as follows.
We assume the \citet[][hereafter: \citetalias{gordon+2003}]{gordon+2003} SMC bar dust extinction law;
even though the true attenuation law may be different at $z=5.7$ than in the local Universe
\citep[e.g.,][]{sanders+2025}, the \citetalias{gordon+2003} still provides a useful
benchmark for comparison with other high-redshift works \citep[e.g.,][]{juodzbalis+2024b,
deugenio+2025d}.
We assume an intrinsic \Halpha/\Hbeta flux ratio of 2.86 for the narrow lines,
and adopt 10 as our fiducial value for the broad lines. The first value is the widely used ratio appropriate for
Case-B recombination and the typical ISM conditions of local galaxies
\citep[e.g.,][]{osterbrock+ferland2006}. The second is motivated by fit quality; we tried models with
intrinsic broad-line decrements of 10, 5, and 3.1 and found that 10 is moderately
preferred over 5, and strongly preferred over 3.1 ($\Delta\,\BIC=3$ and 24, respectively).
We use an erfc prior on \Av too, penalizing solutions with
$\Av>1.8 / 0.44$~mag. This constraint exploits the SED continuum from \cigale, and assumes a constant
scaling between continuum and nebular attenuation. To assess the model response to our assumptions and
priors, we also run otherwise identical setups without the \cigale-based dust prior,
using a separate attenuation for the narrow and broad lines, and adding an \Halpha-only outflow
component; none of these alternative models outperform our default run.
In detail, our \Halpha decomposition is driven purely by
the grating spectrum of \Hbeta and by the model assumptions. As we discuss further, our approach is
driven primarily by the assumptions, but a better model would require medium- or high-resolution
coverage of \Halpha.

The region around \OIIall and \NeIIIL, which includes the Balmer break, is
modelled separately, with a custom continuum model $\mathcal{C}(\lambda)$ given by
\begin{equation}\label{eq.break}
  \mathcal{C}(\lambda) \equiv
  \begin{cases}
      a_0 & \lambda \leq \lambda_\mathrm{break}\\
      A_\mathrm{break} \cdot a_0 + a_1 \cdot (\lambda - \lambda_0) & \lambda > \lambda_\mathrm{break},\\
   \end{cases}
\end{equation}
where $\lambda_0\equiv3728~\AA$ and $\lambda_\mathrm{break} = 3750~\AA$,
while $a_0$, $a_1$, and $A_\mathrm{break}$ are free parameters. We constrain $0\leq A_\mathrm{break} \leq 5$, where $A_\mathrm{break}=1$ would correspond to no discontinuity,
while $A_\mathrm{break} < 1$ or $>1$ would correspond respectively to a Balmer jump or
a Balmer break. The emission lines and the break wavelength $\lambda_\mathrm{break}$
share the same redshift, which is formally a free parameter, but
is strongly constrained by using the posterior probability from the
\Hbeta--\OIIAuall model as prior probability. We also experimented with leaving
$\lambda_\mathrm{break}$ free and found this does not change our conclusions.

\begin{center}
    \begin{figure*}[!t]
        {\phantomsubcaption\label{f.fit.a}
         \phantomsubcaption\label{f.fit.b}
         \phantomsubcaption\label{f.fit.c}
         \phantomsubcaption\label{f.fit.d}
         \phantomsubcaption\label{f.fit.e}
         \phantomsubcaption\label{f.fit.f}
         \phantomsubcaption\label{f.fit.g}
         \phantomsubcaption\label{f.fit.h}
         \phantomsubcaption\label{f.fit.i}
         \phantomsubcaption\label{f.fit.j}
         \phantomsubcaption\label{f.fit.k}
         \phantomsubcaption\label{f.fit.l}}
        \centering
    	\includegraphics[width=\textwidth]{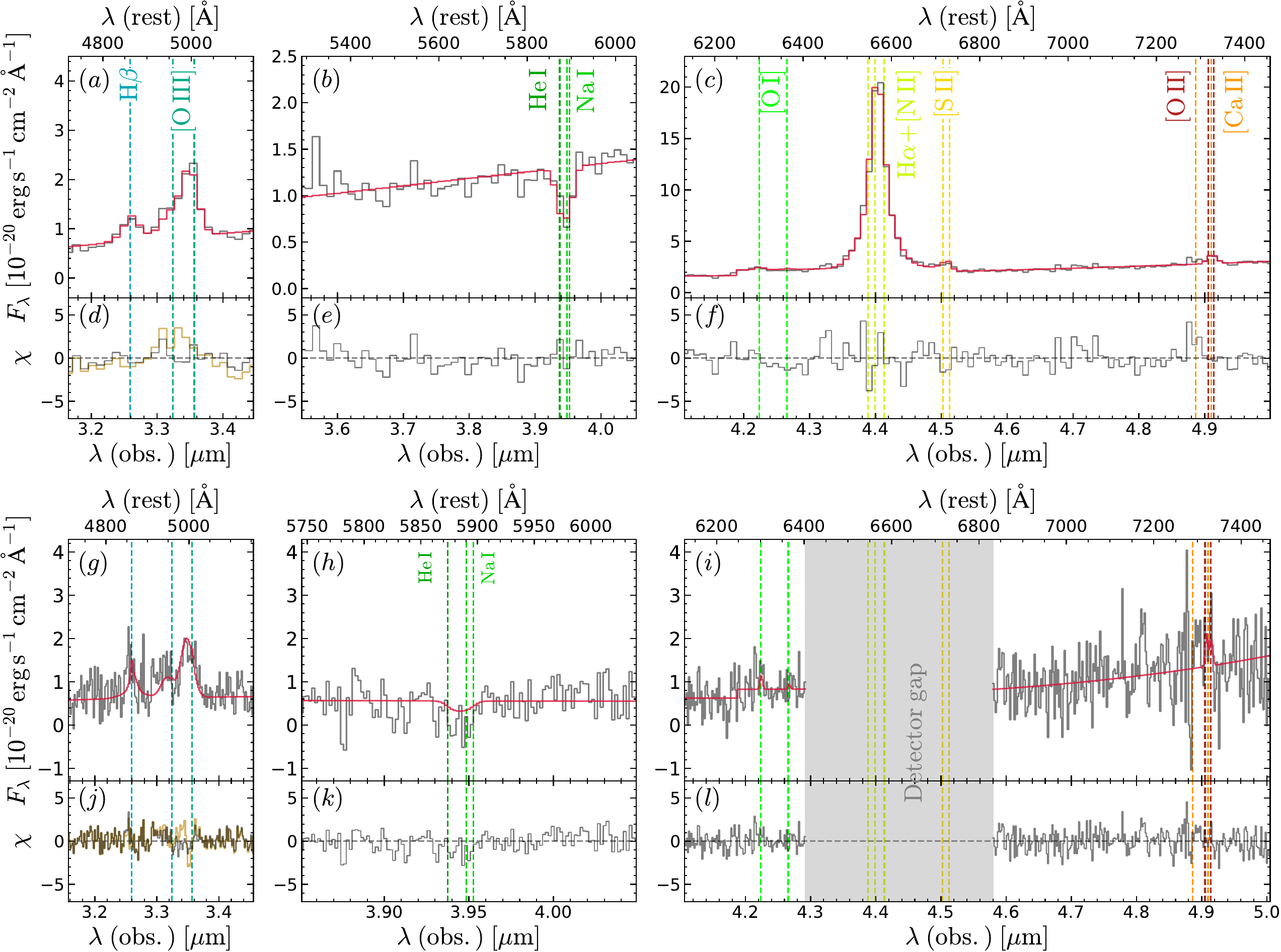}
    	\caption{We model simultaneously data from the prism (panels~a--f) and
        G395M grating (panels~g--l). Panels~a--c and g--i show the data (black)
        and maximum joint-posterior models (red), while panels~d--f and~j--l show the
        $\chi$ residuals. The \Halpha line has a complex profile, which is not
        fully described even by our three-Gaussian model. The sand-coloured
        $\chi$ residuals in panels~\subref{f.fit.d} and \subref{f.fit.j} are from
        a model without an \OIIIall outflow component.
        }\label{f.fit}
    \end{figure*}
\end{center}

The parameters are estimated by comparing this model to both the prism and
grating data simultaneously, with the model being realized on each dataset
using the appropriate line spread function. We use a
2\textsuperscript{nd}-order multiplicative polynomial to account for flux
mismatch between the two dispersers. The resulting physical quantities refer
to the prism spectrum. The model is optimized in a Bayesian framework, using
Markov-chain Monte-Carlo integration, with the setup presented in
\citet{deugenio+2025d}.
The fit quality is shown in Fig.~\ref{f.fit}, where we overlay the data (black) and
maximum joint-posterior model (red).

The results of the spectral fitting are listed in Table~\ref{t.pars} and described in detail in Section~\ref{s.results}.

\section{Spectroscopic results}\label{s.results}

\subsection{Dust attenuation}\label{s.r.ss.dust}

The dust attenuation inferred from the Balmer decrement is highly uncertain: \Halpha falls
in the G395M detector gap, the \Halpha--\NIIall blend is unresolved at prism resolution,
and the \Hbeta and \Hgamma SNRs are too low to obtain meaningful constraints.
For this reason, we compare the family of alternative line models in
Table~\ref{t.modelgrid}. Within the common two-BLR-component, SMC-bar framework, the
best-fitting solutions are those with intrinsic broad-line decrements of 10 and 5.
These return comparatively moderate attenuations, closer to the continuum-based estimate than the lower-decrement alternatives ($A_V\simeq3.6$ and
$4.7$~mag, respectively), and differ very little in fit quality ($\Delta\BIC=3$). A lower
intrinsic decrement of 3.1 \citep[e.g.,][]{dong+2008} requires much larger attenuation,
$A_V\simeq5.5$~mag, and is substantially worse than the fiducial model ($\Delta\BIC\simeq24$).

The control run without the \cigale-based dust prior improves the $\Halpha/\Hbeta=3.1$ model by
$\Delta\BIC\simeq13$ relative to the same model with the prior; this means that the
preference for large decrements is driven in part by the \cigale constraint.
However, even without the prior, the alternative model is still worse than the fiducial
model. In practice, our data do not measure the intrinsic BLR decrement, but they generally
favour models with high observed flux ratio between the \Halpha--\NIIall complex and \Hbeta;
a strongly obscured outflow or high BLR dust could both explain our observations.
However, the more flexible model with separate $A_{V,\mathrm{ISM}}$ and $A_{V,\mathrm{BLR}}$ is
not supported by the data, because it increases the number of parameters without substantially
improving the fit. Likewise, adding an \Halpha outflow or broad \FeII does not improve the fit.
On the other hand, removing the step function $\mathcal{H}$, adopting the Calzetti law,
or forcing a single broad Gaussian are all strongly disfavoured (Table~\ref{t.modelgrid}).

An alternative possibility is very strong \NIIall emission, such as seen in some PSB
galaxies \citep[e.g.,][]{deugenio+2024a}. Since we cannot accurately deblend \NII from \Halpha
given our data, any \NIIall emission beyond our probability prior
(which caps $\NIIall/\SIIall\leq5$, Section~\ref{s.anspec}) would be
interpreted incorrectly as \Halpha, thus artificially increasing $A_V$.

For reference, from \cigale we infer an effective dust attenuation of $A_V = 1.8$~mag,
although the uncertainties of 0.004~mag are clearly under-estimated. While this value
is lower than the line-based estimates of Table~\ref{t.modelgrid}, this is estimated
primarily from the continuum, which dominates the emission in rest-frame $V$-band,
while the corresponding nebular attenuation would be stronger and close to the
findings of the fiducial model.

\subsection{Narrow-line fluxes and SFR}\label{s.r.ss.sfr}

Our narrow-line information relies primarily on the \Hbeta--\OIIIall complex,
which is poorly detected and blended with broad emission, possibly from the BLR
and from ionized outflows (Fig.~\ref{f.fit.g}). The rest of the information
derives from sufficiently isolated lines and doublets, such as \OIIall, \NeIIIL,
\OIL and \SIIall seen in the prism; these lines set an upper limit to the velocity
dispersion, and -- through our informative priors -- also set an upper limit to
the \NIIL/\Halpha ratio.

Given these limitations, we regard the emission-line SFR as very tentative.
In the fiducial model, the dust-corrected narrow-line SFR is $\log(SFR/(\Msun\,\peryr))\simeq1.3$,
while across the range of models tested we reach values of 1.7--2.3 (Table~\ref{t.modelgrid}).
Therefore, the emission-line analysis is consistent with the recently declining SFH inferred by \cigale, but it does not independently establish quiescence, nor can it rule out substantial obscured star formation. Our main evidence for a declining SFH
therefore comes from the SED modelling (Section~\ref{s.i.ss.sed}). Deep, medium-resolution spectroscopy with
simultaneous coverage of \Halpha and \Hbeta is required to break this degeneracy.

\input{table_spec_fit}

\input{table_spectral_models}

\subsection{Broad-line profile and black-hole mass}\label{s.r.ss.blr}

The two broad Gaussians have $FWHM_{\rm b,1}=1200^{+200}_{-100}~\kms$
and $FWHM_{\rm b,2}=3800^{+400}_{-300}~\kms$, with the flux ratio $F_{\rm b,1}/F_{\rm b}
= 0.43^{+0.07}_{-0.05}$. The sum of the two Gaussians has $FWHM_\mathrm{b}=1500^{+200}_{-100}~\kms$.
Using a single Gaussian is disfavoured by the data; repeating the
model inference with a single broad Gaussian, we find a $\Delta\,\BIC>100$.
The measured FWHM value is well beyond the range expected for emission from the
interstellar medium (ISM), but is comparable to what is seen in some fast outflows
in \OIIIL \citep[e.g.,][]{harrison+2014,carniani+2015,husemann+2019,perna+2025b,zamora+2025,bertola+2025}. However, these
outflows are typically seen in massive systems, while more typical outflows
have broadening of 500~\kms \citep{cooper+2025} or less \citep{carniani+2024a}.
Nevertheless, the outflow seen in \OIIIall (Section~\ref{s.r.ss.oution} below) must have a
counterpart in \Halpha too, and this outflow emission can bias the measurement of the
broad-line FWHM. However, our data cannot constrain the detailed properties of the outflow, such as its metallicity, dust attenuation, ionization parameter, and shape of the ionizing field -- all of which should impact the expected \OIIIL/\Halpha ratio. For this reason,
we cannot estimate the outflow contribution to the \Halpha line.
Indeed, dust-obscured outflows have already been detected in the sub-mm galaxies \citep{parlanti+2024}, which also display high-EW broad-line emission (we measure a broad-component EW of $\mathrm{EW(\Halpha)} = -410_{-10}^{+20}~\AA$).

Adding an outflow component with velocity and intrinsic velocity
dispersion tied to the \OIIIall outflow yields a very high \Halpha/\OIIIL flux ratio of
$22\pm5$. This is implausibly high for a blue-shifted outflow, as typical dust-obscured
AGN have outflows with observed ratios of order unity \citep[e.g.,][]{liu+2025}. In our case,
we interpret such a large ratio as due to the SNR and resolution limitations of the data.
The \Halpha outflow appropriates a fraction of the narrowest
broad Gaussian, resulting in lower $F_{\rm b,1}/F_{\rm b}$ and in broader $FWHM_{\rm b,1}$
and $FWHM_{\rm b,2}$, which together yield $FWHM_\mathrm{b} = 2800\pm300~\kms$,
nearly twice as broad as the fiducial value of $FWHM_\mathrm{b} = 1500$~\kms.
A comparison of the BIC value yields $\Delta \BIC = 38$ in favour of the fiducial model without an \Halpha component in the outflow, so we conclude that there is not enough evidence
for a dust-obscured outflow. In our fiducial interpretation, we therefore attribute the broad \Halpha to emission from the
BLR, while noting that unresolved outflow contamination cannot be fully excluded with the current prism data.

Under this fiducial BLR interpretation, we use the measured FWHM and the dust-corrected broad \Halpha flux in the
virial calibration of \citet{reines+volonteri2015}, and find a black-hole mass
$\log(\mbh/\Msun) = 8.1\pm0.7$. This value should be regarded as an illustrative virial estimate rather than a secure black-hole mass, because the \Halpha--\NIIall blend is unresolved and the separation between BLR and outflow emission is model dependent. The uncertainties are dominated by model assumptions (0.4 dex) and by the scatter in the calibration
\citep[0.55 dex;][]{reines+volonteri2015}, for a total of 0.7~dex, after summing in quadrature.
To estimate the systematics of 0.4 dex, we reasoned that increasing the very uncertain dust attenuation from $A_V \sim 3$~mag to 6~mag would increase \mbh by 0.4~dex. A similar
change is obtained when including the outflow component of \Halpha in the model, which
increases \mbh by a factor of 2.6 (the twice broader FWHM we have seen in the previous paragraph is compensated by lower line flux). For this reason, our subsequent interpretation does not rely on the precise value of \mbh, but only on the robust multi-wavelength evidence for a luminous AGN.

To estimate the bolometric luminosity we use the calibration of \citet{stern+laor2012}, based
on the broad \Halpha line. We use the dust-corrected luminosity
$\log(L_\mathrm{b}(\Halpha)/(10^{42}~\ergs)) = 2.54^{+0.05}_{-0.07}$ and obtain
$\log(L_\mathrm{bol}/(\ergs)) = 46.65^{+0.05}_{-0.07}$, where the uncertainty is
formal only and does not include the $\simeq0.5$-dex systematics of the calibration.
This is an order of magnitude
brighter than most luminous LRDs \citep[e.g.,][]{juodzbalis+2024b,loiacono+2025},
with the only exception of \monster \citep{greene+2024,labbe+2024}, which has
$\log(\Lbol/(\ergs)) = 46.3\pm0.2$ \citep{matthee+2024b}. Our broad-\Halpha estimate
is $0.5$~dex higher than the fiducial \cigale estimate of $\log(\Lbol/(\ergs))=46.18$
(Section~\ref{s.i.ss.sed}); the two are consistent within the systematic uncertainties
of both methods (1--2~\textsigma; Section~\ref{s.i.ss.sed}).

\subsection{Balmer break}\label{s.r.ss.bbreak}

\begin{center}
    \begin{figure}[!t]
        {\phantomsubcaption\label{f.bbreak.a}
         \phantomsubcaption\label{f.bbreak.b}}
        \centering
    	\includegraphics[width=\columnwidth]{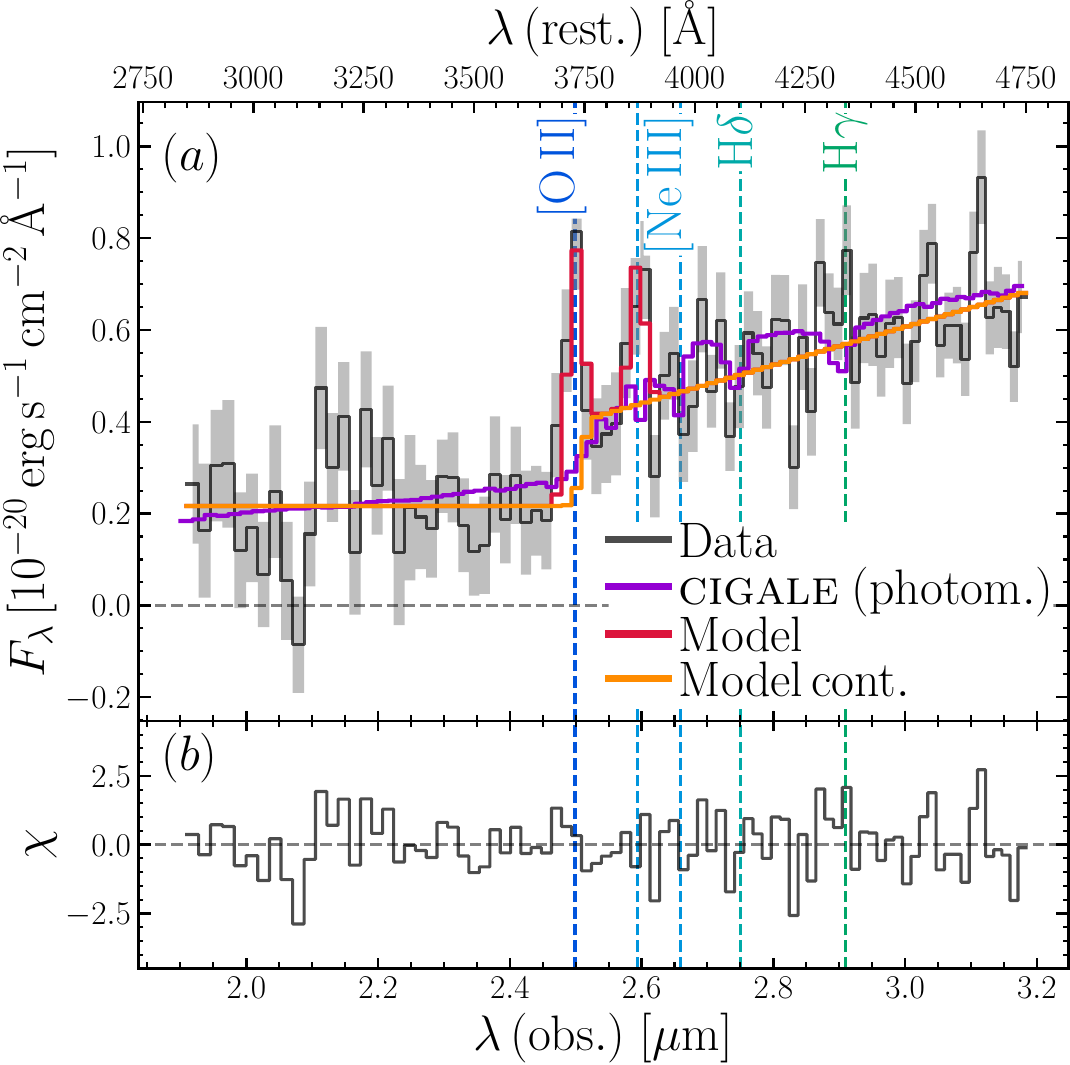}
    	\caption{a. Spectral region near the Balmer limit, illustrating the data
        and maximum joint-posterior model. The model continuum is shown in orange; it
        coincides with the maximum joint-posterior model everywhere, except near the
        emission lines. The presence of a spectral break is clear both in the data
        and in the model (4-\textsigma significance), but we lack the spectral
        resolution and SNR to characterize the exact shape of the break. The purple line is the best-fit \cigale model from Fig.~\ref{f.cigale.c}, which reproduces the break reasonably well, even though
        \cigale is not fit to the spectroscopy.
        b. $\chi$ residuals, highlighting the good fit.
        }\label{f.bbreak}
    \end{figure}
\end{center}

The spectral region near the Balmer limit is displayed in Fig.~\ref{f.bbreak}. A Balmer
break is visible in the data, and is qualitatively consistent with the best-fit model from \cigale (downscaled to match the continuum level, and overlaid in purple in Fig.~\ref{f.bbreak.a}).
Statistically, the \cigale model represents an adequate match to the data, with a reduced $\chi^2_\nu = 1.7\text{--}2.2$ (where the range
is due to subtracting no or all free parameters from the number of degrees of freedom, since we do not optimize \cigale on the spectrum,
but on the photometry; see Section~\ref{s.i.ss.sed}).

The presence of a Balmer break is independently confirmed by a statistical analysis of the spectrum.
First, replacing the fiducial continuum model of Eq.~\ref{eq.break} with a
2\textsuperscript{nd}-order polynomial and with no break yields a worse fit, with $\Delta \BIC = 11$ in favour of the fiducial model. If we extend the fiducial model by letting
the break wavelength $\lambda_\mathrm{break}$ vary freely between 3600 and
4000~\AA, the resulting posterior distributions give $A_\mathrm{break}=1.8\pm0.2$ and $\lambda_\mathrm{break} =
3740_{-70}^{+50}~\AA$, which is fully consistent with the effective wavelength of the Balmer break.
From the MCMC chains, we calculate an emission-line corrected $D_\mathrm{n}4000$
index of $1.18_{-0.02}^{+0.03}$ \citep{balogh+1999} and a Balmer break index of
$3.5_{-0.3}^{+0.4}$ \citetext{where we have used the standard wavelength windows
of \citealp{curtis-lake+2023}}. Note that the Balmer-break index is more sensitive
to dust attenuation than $D_\mathrm{n}4000$.

Statistically, the Balmer break can be established by calculating the probability
$P(A_\mathrm{break}\geq1)$ in the fiducial model. After ensuring that our
sampler contains more than $10^7$ points, we can
calculate the probability that $A_\mathrm{break}\geq1$ directly from the posterior.
This yields $P(A_\mathrm{break}\leq1)<2\times10^{-6}$, which would correspond to a
one-tailed significance of better than 4.5 \textsigma.

If we repeat the analysis using the CAPERS spectrum, we infer $P(A_\mathrm{break}\leq1)<10^{-7}$, i.e. better than 5~\textsigma significance, limited by the number of chains for calculating the posterior
probability. Interestingly, CAPERS finds a weaker break amplitude than RUBIES, $A_\mathrm{break} = 1.49\pm0.05$.
This difference is likely due to spatial variations in the SED, since
the (nominal) position of the CAPERS micro-shutter is off-centre.
Confirming this hypothesis would require spatially resolved spectroscopy.

In any case, since a discontinuity consistent with the Balmer break is supported independently by RUBIES and CAPERS, and by the
photometry (Fig.~\ref{f.cigale}) and SED modelling (Section~\ref{s.i.ss.sed}), we conclude that \target does possess a
Balmer break.

Our $D_\mathrm{n}4000$ measurement can be used to infer directly a simple stellar population (SSP) age. By comparing to the \citet{bruzual+charlot2003} models \citep[assuming a single burst with solar or 0.2-solar metallicity; e.g.,][]{kauffman+2003} we would infer a burst age
of 250~Myr for $D_\mathrm{n}4000=1.18$, while
$D_\mathrm{n}4000=1.00$ (corresponding to a 4-\textsigma
deviation with respect to our measurement) would still
imply an age older than 50~Myr. So the strength of the spectral break
would correspond to intermediate SSP ages between star-formers and post-starburst galaxies.

\renewcommand{\arraystretch}{1.5}
\setlength\extrarowheight{1pt}
\begin{table}
    \caption{Free parameters of the model describing the spectral region around the
    Balmer limit (see Fig.~\ref{f.bbreak}).
    $^\ddag$ The break is detected regardless of whether $\lambda_\mathrm{break}$ is fixed or free.
    }
    \centering
    \begin{tabular}{cccc} 
        \hline
        Parameter & Prior & Unit & Posterior \\
        \hline
        \hspace{-0.3cm}$z_\mathrm{n}$           & \hspace{-0.3cm}$\mathcal{G}(5.702, 0.001)$ & ---  & --- \\
        \hspace{-0.3cm}\sign                    & \hspace{-0.3cm}$\mathcal{G}(90, 100)$      & \kms & --- \\
        \hspace{-0.3cm}$F(\OIIall)$             & \hspace{-0.3cm}$\mathcal{U}(0, 10)$        & \hspace{-0.5cm}\fluxcgs[\!-\!18][] & $1.5^{+0.5}_{-0.6}$ \\
        \hspace{-0.3cm}$F(\NeIIIL)$             & \hspace{-0.3cm}$\mathcal{U}(0, 10)$        & \hspace{-0.5cm}\fluxcgs[\!-\!18][] & $0.8^{+0.3}_{-0.3}$\\
        \hspace{-0.3cm}$\lambda_\mathrm{break}^\ddag$ & \hspace{-0.3cm}$\mathcal{U}(3600, 4000)$   & \AA           & $3740^{+50}_{-70}$ \\
        \hspace{-0.3cm}$A_\mathrm{break}$       & \hspace{-0.3cm}$\mathcal{U}(0, 5)$        & ---           & $1.8^{+0.2}_{-0.2}$ \\
        \hline 
    \end{tabular} 
    \label{t.bbreak}
\end{table}

\subsection{Ionized outflow}\label{s.r.ss.oution}

The need for an \OIIIall outflow component is visible in the fit residuals; when
the outflow component is removed, the best-fit model leaves structured positive
residuals blueward of \OIIIL, in both the prism and independently in the G395M
spectra (sand lines in Fig.~\ref{f.fit.d} and~\subref{f.fit.j}). The fiducial model
captures this excess with a broad, blueshifted \OIIIall component, consistent
with the complex \OIIIall profile seen in the G395M spectrum
(Fig.~\ref{f.data.c}), although the grating spectrum has low SNR in this region.
Crucially, the improvement in the residuals is statistically favoured even after
penalizing the three additional outflow parameters, since we find $\Delta \BIC = 22$
in favour of the model with outflows.

Although the parameters are very
uncertain, we can attempt to estimate a mass outflow rate, following the standard
methodology of \citet{carniani+2015}, which has been applied to both AGN \citep[e.g.,][]{kakkad+2022} and star-forming galaxies \citep[e.g.,][]{carniani+2024a}. We adapt the formula
reported in \citet{deugenio+2024a}
\begin{equation}
    \begin{split}
    \Mout &= 8\times10^7 \cdot \left( \dfrac{1}{10^{\log (\mathrm{O/H}) - \log (\mathrm{O/H})_\odot}} \right) \\
    &\hspace{1cm} \cdot \left( \dfrac{L_{\OIIIL}}{10^{44}\,\ergs} \right) \left(\dfrac{\nelec}{500\,\pcm} \right)^{-1}\,\Msun\\
    \Mdotout[{\OIII}] &= 3 \, \left(\vert v_\mathrm{out} \vert + 2 \, \sigma_\mathrm{out}\right) \, \frac{M_\mathrm{out}}{\Rout},
    \end{split}
\label{eq.o3out}
\end{equation}
where $\mathrm{O/H}$ is the outflow metallicity, expressed by the abundance ratio of oxygen to hydrogen atoms, $L_{\OIIIL}$ is the outflow luminosity (without dust attenuation correction, lacking any constraints), \nelec is the electron density, $v_\mathrm{out}$ and $\sigma_\mathrm{out}$ refer to the model parameters of the \OIIIall outflow, and \Rout is the extent of the outflow. Compared to \citet{deugenio+2024a}, we replaced their spatially and spectrally resolved $v_\mathrm{out}$ with $\vert v_\mathrm{out} \vert + 2 \sigma_\mathrm{out}$
\citep[e.g.,][]{rupke+2005}.
For lack of constraints on the physical properties of the ionized gas in the outflow,
we adopt solar metallicity, as massive, dusty galaxies tend to be metal rich. We set $\nelec = 2200~\pcm$, equal to the geometric mean
of the range of electron densities observed in outflows \citep[$\nelec = 500\text{--}10,000~\pcm$;][]{baron+netzer2019,davies+2020}. Finally, we use $\Rout = 1$~kpc, set by
the size of the UV-bright clump. We calculate the full posterior distribution of
\Mdotout[{\OIII}] from the MCMC sampler, obtaining $\log(\Mdotout[{\OIII}]/(\Msun\,\peryr)) = 0.36\pm0.04$, but we stress that the systematic uncertainties are much larger. Varying
\nelec between 500 and 10,000~\pcm, \Mdotout[{\OIII}] spans 1.3~dex, or a factor 4.5
additional uncertainty.

\subsection{Neutral gas absorption and possible outflow}\label{s.r.ss.outnai}

The presence of \NaIall absorption is extremely clear in the prism spectrum
(Fig.~\ref{f.data.b}) and, tentatively, in the G395M spectrum (Fig.~\ref{f.data.c}).
Our joint analysis yields an $\mathrm{EW(\NaIall)} = 17\pm2 \AA$ -- much higher than any stellar-population model, even the most metal-rich local galaxies with the highest Na/Fe abundance
ratio \citep[$EW(\NaIall)\lesssim 6~\AA$ for $\mathrm{[Na/Fe]=0.5\text{--}0.7~dex}$,
][]{labarbera+2016}. High $EW$ absorption is also confirmed by the CAPERS
spectrum, where we measure $15\pm1$~\AA (Appendix~\ref{a.metal}).
Our data cannot establish if the absorption
is observed in the rest frame \citetext{as seen in many hydrogen absorbers in
LRDs; \citealp{deugenio+2025d,ma+2025,deugenio+2025e}}, or if it is blue-shifted.
This is because we find a velocity $v_\mathrm{abs} = -400^{+200}_{-100}~\kms$,
which is only 2-\textsigma away from rest (defined as $v = 0$~\kms). The absorber
velocity dispersion is $\sigma_\mathrm{abs} = 300^{+100}_{-100}~\kms$, which again
comes with too large uncertainties to assess if the line is intrinsically
broad or not.

Under the outflow hypothesis -- strengthened by the detection of the \OIII outflow -- we can estimate the mass outflow rate using the methods
outlined in \citet{rupke+2005}. From their formulae, and using the same assumptions about
the gas metallicity and ionization state as \citet{davies+2024}, we can calculate the
mass outflow rate for each sample in the set of Markov chains, and we obtain
$\log(\Mdotout[{\NaI}]/(\Msun\,\peryr)) = 2.3^{+0.2}_{-0.3}$. This value is remarkably
large, among the highest values reported e.g. in massive galaxies at
$z=1\text{--}2$ \citep{davies+2024,sun+2025b}.
The reported precision and the statistical significance of the outflow may seem
surprising, given the large uncertainties on $v_\mathrm{abs}$ and $\sigma_\mathrm{abs}$.
However, having defined the outflow velocity as $v_\mathrm{out} \equiv \vert
v_\mathrm{abs} \vert + 2\cdot \sigma_\mathrm{abs}$, we obtain a much larger $v_\mathrm{out}$ value than $\vert v_\mathrm{abs} \vert$ alone, one that is clearly
inconsistent with 0 (unlike $v_\mathrm{abs}$ itself). This means that, under the outflow
hypothesis, the data suggest a high mass outflow rate, yet the current SNR is insufficient
to establish if outflows are the correct scenario.

\section{A regular X-ray emitter}\label{s.xray}

We derive the X-ray luminosity directly from the published catalogue of \citet{nandra+2015}.
To estimate the luminosity in the 2--10~keV range, we assumed a photon index $\Gamma = 1.9$,
which is the median value for normal AGNs at this redshift \citep{schemmer+2006,just+2007,nanni+2017,vito+2019}.
We obtain a range of $L_\mathrm{2\text{--}10\,keV}=(2.3\text{--}3.8) \times 10^{44}~\ergs$,
where the limits of the interval correspond to the extreme cases of no absorption
($N_H = 0~\pcm[2]$) or Compton-thick absorption ($N_\mathrm{H} = 10^{24}~\pcm[2]$).

\begin{center}
  \begin{figure}[!t]
    \centering
    \includegraphics[width=\columnwidth]{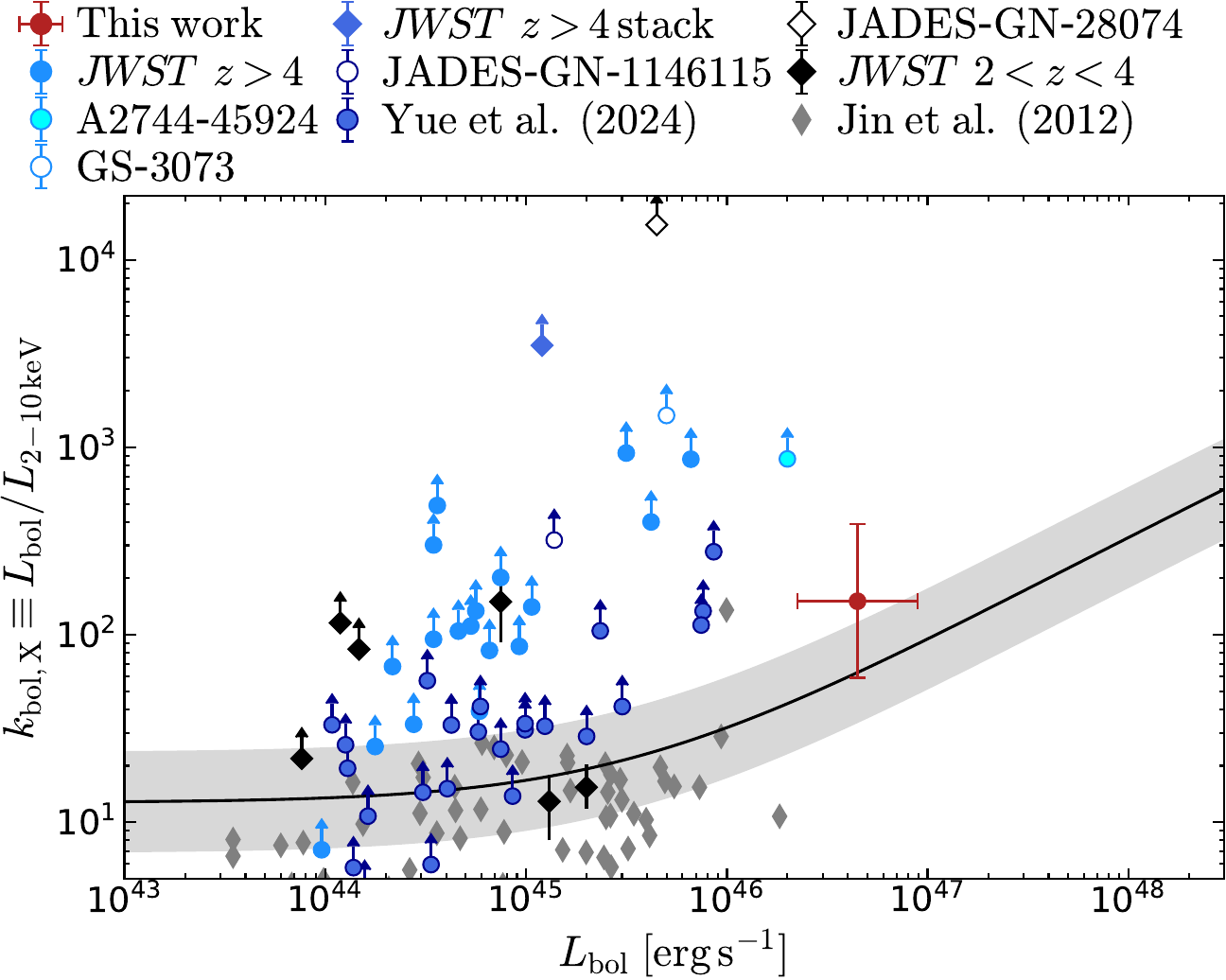}
    \caption{Ratio \kbolX between AGN bolometric luminosity and the X-ray
      (2--10 keV) luminosity, as a function of bolometric luminosity. \target
      (red circle) stands out compared to other \jwst discovered AGN and LRDs at
      $z>2$ (circles and diamonds) in lying much closer to the local
      \Lbol--\kbolX relation \citep[black line with grey shaded region;][]{
      duras+2020} than other \jwst AGN, consistent with it within the uncertainties.
      In our fiducial model, \target is more luminous than all known LRDs,
      including both individual measurements \citetext{light and dark blue,
      \citealp{maiolino+2025x,yue+2024}; and their stacks, blue diamond,
      \citealp{maiolino+2025x}}. We also indicate local Seyfert 1 \citep[grey
      diamonds;][]{jin+2012}. While the most luminous LRDs \citep[e.g.,
      \monster;][]{labbe+2024,matthee+2024b} come close to \target, they all
      remain undetected. Hence X-ray detection in \target is not merely a
      consequence of its high \Lbol.
    }\label{f.xrays}
  \end{figure}
\end{center}

In Fig.~\ref{f.xrays} we show \kbolX, the ratio between \Lbol and
$L_\mathrm{2\text{--}10\,keV}$,
as a function of \Lbol. By comparing \target to quasars and other \jwst AGN, we
infer that \target is not X-ray weak, unlike most \jwst-discovered AGN
at $z>2$ \citep{yue+2024,maiolino+2025x}, which include some of the most luminous
LRDs known \citep{juodzbalis+2024b,wang+2025,loiacono+2025}, and even blue, luminous
AGNs \citep{ubler+2023}. While \target is 10 times brighter than typical high-redshift AGNs
-- making detection in X-rays easier -- several LRDs and high-redshift AGNs have
\textit{more stringent} constraints on \kbolX than \target, due to being at lower
redshifts \citep[e.g.,][]{juodzbalis+2024b,loiacono+2025}, or in deeper \textit{Chandra}
fields \citep[e.g.,][]{ubler+2023}, or both.

\section{Discussion}\label{s.disc}

\subsection{The Physical Nature of \target: A Massive, Quenching Galaxy with Powerful Outflows}

\target is massive, $\mstar \sim 10^{10.5\text{--}11}~\Msun$, as implied by its luminosity and by the
pronounced, spatially resolved Balmer break and inferred from the SED fit
(Section~\ref{s.i.ss.sed}). For reference, these values are larger than the `knee'
of the galaxy mass function at $z=5\text{--}6$,
\citep[$10^{10\text{--}10.3}~\Msun$;][]{weaver+2023,weibel+2024}.
In terms of mass formed, \target is already on par with massive quiescent
galaxies at redshifts $z\sim5$ \citetext{e.g., GS-9209, \citealp{carnall+2023}; RUBIES-QG-1, \citealp{degraaff+2025b}}.

The source is compact yet resolved in all robust NIRCam detections
(Section~\ref{s.i.ss.morph}).
The size decreases with increasing wavelength (Fig.~\ref{f.sizes} and
Table~\ref{t.sizes}), a fact that is consistent with either an increasing contribution
from a point source at red wavelengths or centrally concentrated dust. Using
the size in F356W (rest-frame $\sim$0.5~\mum) as reference, we find
$\re \simeq 380$~pc, fully consistent with the value $\re(z=5.7) = 350$~pc,
predicted from the redshift-dependent relation of quiescent galaxies of comparable
stellar mass \citep{ji+2024a}. This size match makes \target a plausible progenitor,
without requiring substantial kinematic transformation.

There is an apparent discrepancy in the size of \target, whose rest-frame $V$-band \re is 2--3
times larger than for e.g., GS-9209 \citetext{$\re = 400\text{--}600$~pc,
Section~\ref{s.i.ss.morph}; vs $\re=200\text{--}250$~pc, \citealp{carnall+2023,
ji+2024a,pascalau+2025}}. However, as we noted, the half-light radius of \target
may be affected by the centrally concentrated dust, hence the
size mismatch with GS-9209 is only apparent, not structural; eventually, dust removal
would bring the V-band \re of \target towards that of GS-9209.
In fact, the two sizes are already comparable in F410M ($\re = 210$~pc), a band
that is not affected by \Halpha.
In agreement with the concentrated dust hypothesis, the continuum slope becomes
less steep going from the centre (probed by RUBIES, Fig.~\ref{f.data.a}
and~\ref{f.data.b}) to the outskirts (probed by CAPERS). We should be cautious,
however, since potential contribution from a point-source continuum
at redder wavelengths cannot be ruled out by our data.

The SED-inferred SFH shows a recent declining trend in SFR with time, consistent with quenching
already underway. The SFR averaged over 100 Myr is almost 50~percent higher than the
value over the last 10~Myr ($\log(SFR_\mathrm{100\,Myr}/(\Msun\,\peryr)) = 1.46\pm0.02$ and
$\log(SFR_\mathrm{10\,Myr}/(\Msun\,\peryr)) = 1.30\pm0.04$, respectively).
Admittedly, this conclusion is highly uncertain, and relies on the \cigale assumptions
for nebular emission, dust properties, energy balance, and SFH priors.
While FIR emission can be used to infer the current SFR, cold dust in quasar-host
galaxies can also be heated by the AGN \citep{kirkpatrick+2015}, which adds to the
uncertainties. While the exact role of quasar heating is still debated \citep{stanley+2018}, intense FIR emission can be present even in galaxies with little
ongoing star formation -- particularly in recently quenched galaxies.
Here dust
heating can be powered by intense, non-ionizing UV emission from evolved young stars
\citep{wu+2022,deugenio+2024a}.
With all these uncertainties,
we compare the SED-based inference to the emission-line modelling, which gives a fiducial SFR of $\log(SFR/(\Msun\,\peryr)) \simeq 1.3$
(Section~\ref{s.r.ss.sfr}) for the last 3--10~Myr. This measurement however suffers from substantial uncertainties.
To start, it neglects any contribution from the AGN narrow-line region and outflows (which would
drive the SFR down). In addition, it relies on our modelling of the broad lines and possible outflow
components, which in the prism are degenerate with the narrow lines.
If we allow for much higher $\NIIL/\SIIall>5$, this would increase the \NIIall contribution to the
unresolved \Halpha--\NIIall blend, and result in a lower SFR still. Admittedly, a precise
characterization of the recent SFH requires better data than what is currently available.

\begin{figure}[!t]
  {\phantomsubcaption\label{f.sfhcomp.a}
   \phantomsubcaption\label{f.sfhcomp.b}}
  \centering
  \includegraphics[width=\columnwidth]{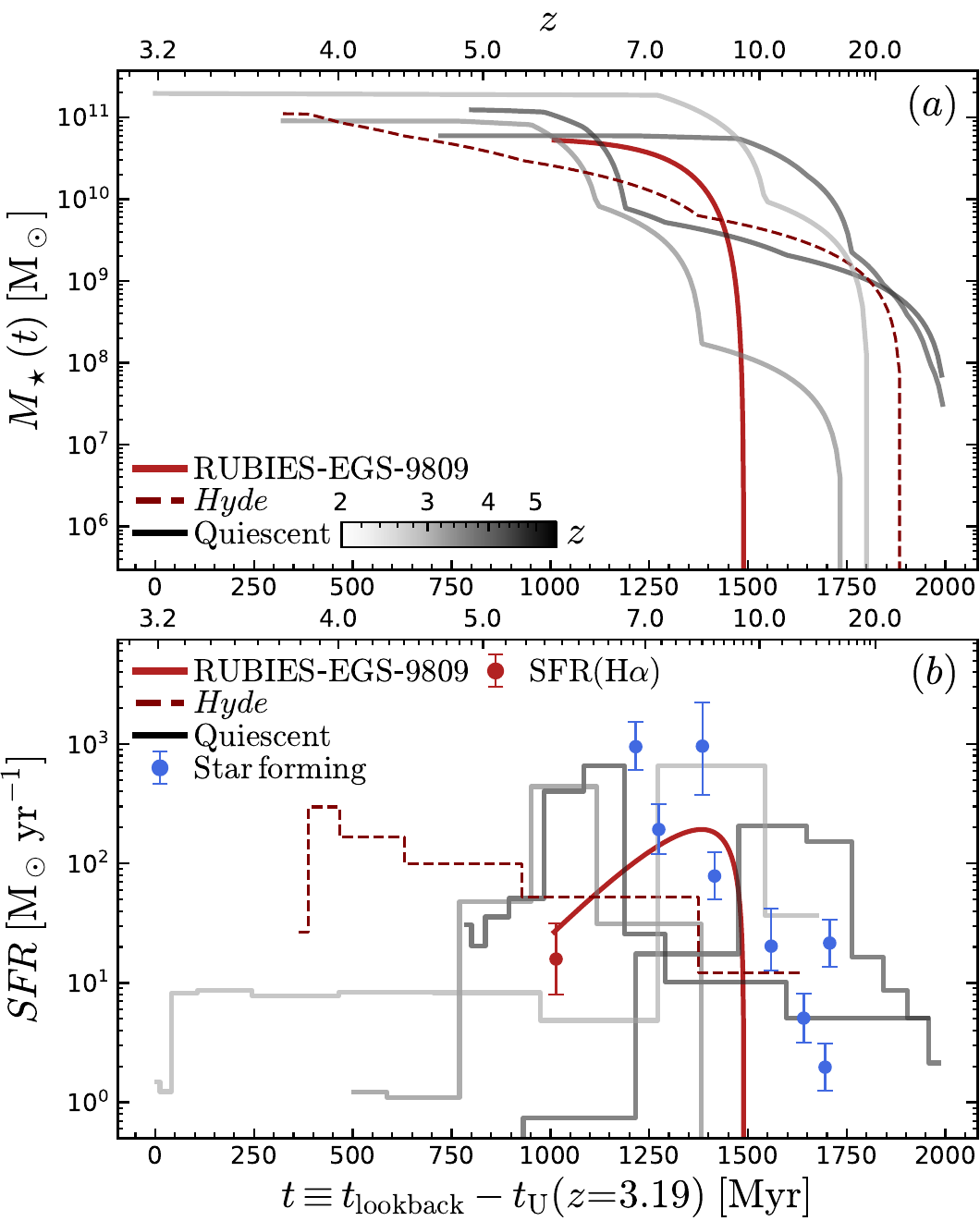}
  \caption{The SFH of \target is similar to that of massive, quiescent
    galaxies at $z\gtrsim3$. Panel~\subref{f.sfhcomp.a} shows the
    cumulative SFH, highlighting the similarity in final \mstar.
    Greyscale curves are massive, quiescent galaxies from the literature
    \citetext{in order of increasing redshift: \citealp{glazebrook+2024},
    with the SFH from \citealp{turner+2025}; \citealp{glazebrook+2017}, with
    the SFH from \citealp{perez-gonzalez+2025}; \citealp{carnall+2023}, with
    the SFH from \citealp{ji+2025b}; and \citealp{degraaff+2025b}}. The
    red dashed curve is the compact and dusty quenching galaxy \textit{Hyde}
    \citetext{\citealp{schreiber+2018}, with the SFH from
    \citealp{perez-gonzalez+2025}}.
    Panel~\subref{f.sfhcomp.b} shows the SFR vs look-back time; we also show
    the 10-Myr SFR from \Halpha, and a set of SFR measurements from the
    literature, compiled by \citet{turner+2025}. All look-back times are
    matched to the lowest-redshift object, at $z=3.19$.
   }\label{f.sfhcomp}
\end{figure}

This current SFR -- whether inferred from \cigale or from the emission lines --
places our galaxy below the star-forming main sequence (SFMS). However, the
exact distance from the SFMS depends on which scaling relation we use, ranging from a
factor of 3 to over one dex below \citep{speagle+2014,popesso+2023,cole+2024,
clarke+2024}. Since most definitions of quiescence require at least one dex below
the SFMS for 100~Myr, we cannot confirm or rule out if \target is quiescent.
The decline from the 100-Myr-averaged SFR to the 10-Myr-averaged SFR (Section~\ref{s.i.ss.sed})
therefore suggests that \target is a strong quenching candidate.
As an order-of-magnitude consistency check, the SCUBA-2 flux density ($S_{850\,\mum}=0.48\pm0.12$~mJy) corresponds to an obscured SFR of order a few $\times 10~\Msun~\mathrm{yr}^{-1}$ \citep[e.g.,][]{kennicutt+evans2012}, which is broadly consistent with a recently declining SFR, and disfavours hundreds of $\Msun~\mathrm{yr}^{-1}$, as inferred from the peak of the SFH (e.g., Fig.~\ref{f.sfhcomp.b}).

In Fig.~\ref{f.sfhcomp} we compare our target to a set of massive galaxies at
$z=3\text{--}5$ \citep[e.g.,][]{carnall+2023,glazebrook+2024}. The presence of a
Balmer break implies an already evolved stellar population, and the tentative downturn
in the SFR (as inferred from \cigale, Section~\ref{s.i.ss.sed}) suggests that the
galaxy may be in the process of quenching. We detect multiphase outflows: broad
\OIII consistent with an ionized outflow ($\log(\Mdotout[{\OIII}]/(\Msun\,\peryr)) =
0.36\pm0.04$) and deep \NaI absorption whose large EW requires an ISM origin.
Interpreted as an outflow, this implies a high neutral-phase mass-outflow rate
($\log(\Mdotout[{\NaI}]/(\Msun\,\peryr)) = 2.3^{+0.2}_{-0.3}$). This is another case of multiphase outflows
at high redshift in which the mass outflow rate is higher in the neutral phase than in the ionized one \citep{davies+2024,deugenio+2024a,belli+2024,parlanti+2025}.
While both outflow rates have large systematics, their coincidence with a luminous AGN
strongly favours AGN-driven clearing of the ISM.

The duration of this phase cannot be derived solely from our data, but we can
provide a rough estimate by comparing the expected gas content to the mass
outflow rate.
For a galaxy with $\mstar = 10^{10.5\text{--}11}~\Msun$ on the star-forming
sequence, we expect a molecular-to-stellar mass ratio of $\sim 75$~percent
\citep{tacconi+2020}. This must be an upper limit, since the downturn in SFR
suggests that quenching has been already underway.
Dividing the inferred molecular gas mass by the mass rate of the neutral-gas
outflow, we obtain a depletion time $\tau_\mathrm{dep} \lesssim
150\text{--}450$~Myr. This is clearly too long to explain the lack of
progenitors for massive, quiescent galaxies at $z\gtrsim3$. Moreover, we have
not even considered the fact that over such a long time period, main-sequence
galaxies are expected to further accrete substantial amounts of gas from the
cosmic web \citep{tacconi+2020}.
A shorter depletion time is possible if the galaxy has a low molecular gas
fraction, and if the system is no longer undergoing cold-gas accretion, for
instance due to preventative feedback being already at work
\citep[e.g.,][]{scholtz+2024}. Alternatively, the
mass outflow rate could be under-estimated, for instance if our assumptions
about the metallicity and ionization state of the neutral gas are incorrect,
or if we are missing entirely an even colder phase of the outflow \citep[which
is seen in high-redshift galaxies; e.g.,][]{herrera-camus+2019,jones+2019}.

Deeper spectroscopy can help determine more precisely the
recent SFH of this galaxy, thus providing independent constraints on the
quenching timescale \citep[e.g.,][]{carnall+2023,park+2024,baker+2025}.

To provide an order-of-magnitude reference estimate for the space density of
sources like \target, we rely on its X-ray detection. At $z=5.7$, \target is the
only dust-obscured X-ray source \citep{nandra+2015} in the CEERS survey footprint
of 90~arcmin$^2$ \citep{finkelstein+2025}.
To calculate the survey volume, we consider redshift limits of $z\geq3\text{--}4$
\citep[since this is the redshift range where the abundance of massive quiescent
galaxies is in tension with theory;][]{carnall+2023,valentino+2023,baker+2025}
and $z\leq7.3$ (since above this redshift \Halpha is not observable by NIRSpec).
With these limits, and the assumed cosmology, we find a comoving number density
$n = (0.9\text{--}1.3)\times10^{-6}~\mathrm{cMpc}^{-3}$. Clearly, this estimate
cannot be interpreted as the total number density of dusty, quenching
galaxies, nor as a lower limit on the cosmic mean. First, it is based on a single
source, hence it is subject to large Poisson uncertainty. Second, by relying on a
single field, we are subject to cosmic variance \citep[e.g.,][]{valentino+2023}.
Finally, this is conditional on an X-ray selection, and is therefore incomplete
for X-ray faint AGN, including intrinsically X-ray-weak AGNs, heavily absorbed
sources, and systems in which the AGN has already faded. In particular, we are
certainly missing objects that are more naturally selected through deep MIR
observations.

With these limitations in mind, the CEERS X-ray-selected value is ten times
lower than the number density of massive, quiescent galaxies, which is around $n =
10^{-5}~\mathrm{cMpc}^{-3}$. The two numbers could be reconciled if the visibility
of the dusty-progenitor phase was significantly shorter than that of massive
galaxies. This shorter visibility could arise if -- for example -- the breakout phase
is short compared to both the dust-obscured starburst and the quiescent phase,
or, alternatively, from the limitations outlined above.
This is particularly relevant because the other compelling candidate, the dusty
post-starburst galaxy \textit{Hyde}, is undetected in the X-rays \citetext{but
the relevant X-ray data are much shallower than for \target; cf.~\citealp{
nandra+2015} vs \citealp{civano+2016}}.
The purpose of this work is therefore not to provide a definitive measurement
of the number density of dusty, quenching AGN hosts. Rather, \target demonstrates
that such systems exist, and that they can combine an evolved stellar population,
heavy dust obscuration, luminous AGN activity, and possible outflows at $z>5$.
Establishing whether this pathway is common enough to contribute substantially
to the progenitor population of massive quiescent galaxies will require a
dedicated census across multiple fields, with joint X-ray, MIR, FIR, radio, and
\jwst selection functions.

\subsection{A massive galaxy, not a Little Red Dot AGN}\label{s.d.ss.notlrd}

\jwst has discovered a new population of broad-line AGNs, whose unique characteristics
are extremely rare in the local Universe \citetext{\citealp{ma+2025}, \citealp{lin+2025b},
\citealp{bisigello+2025}, \citealp{lin+2025b}}. These include systems in which
even the optical continuum is AGN-dominated \citep{ji+2025a,lin+2025b,
deugenio+2025g}. These works cast doubt on our ability to accurately measure
\mstar in LRDs \citep{wang+2025,juodzbalis+2024b,ma+2025b}. In fact, strong Balmer breaks in
LRDs \citep{furtak+2024,labbe+2024,wang+2024b} are demonstrably associated with low-\mstar
galaxies, as inferred from both their clustering \citep{pizzati+2025,matthee+2024b,lin+2025c}
and from the physical upper limit set by their
low dynamical masses \citep{wang+2025,juodzbalis+2024b,ji+2025a,deugenio+2025d,akins+2025}.
Recently, a possible LRD embedded in an extended galaxy has also been discovered
\citep{rinaldi+2025b}.

Since our source \target presents several characteristics in common with LRDs, it is worth
assessing whether it may be an LRD itself, or at least an intermediate system, similar to
\textit{Saguaro} \citep{rinaldi+2025b}. The answer to this question is consequential for our
confidence in the inferred \mstar and SFH.
The overestimated \mstar in LRDs stems from the incorrect interpretation of the
Balmer break, which in these AGNs arises from dense-gas absorption near the black hole, and not in
the atmosphere of evolved stellar populations \citep{ji+2025a,naidu+2025,degraaff+2025}.

Two diagnostics argue against the LRD interpretation. First, the emission redward of the
Balmer break is spatially extended: we measure $\re\sim0.5$~kpc in F277W (which samples the
continuum just longward of the break) and a similarly extended size in F200W
(Section~\ref{s.i.ss.morph}).
Although F277W includes \OIIall and \NeIIIL, these lines contribute only
$\simeq3$~percent of the F277W broad-band flux when the fitted line fluxes are
convolved with the filter curve (Section~\ref{s.i.ss.morph}). Thus the extended
F277W morphology cannot be explained primarily by extended ionized gas.
Combined with the strong colour contrast across the break (F277W/F200W$\simeq4.5$), this
implies that the Balmer-break continuum is not dominated by a compact, unresolved component.
The measured size is an order of magnitude larger than in LRDs, where the Balmer break is always
extremely compact, and almost never spatially resolved
\citetext{e.g., \citealp{furtak+2024} place upper limits of $\re<30$~pc, while
\citealp{degraaff+2025} measure $\re = 40$~pc}.

Second, \target is clearly detected in X-rays, MIR and FIR wavelengths. In contrast,
spectroscopically confirmed LRDs are generally deficient in all
three wavelength ranges \citep{ananna+2024,yue+2024,maiolino+2025x,kokubo+harikane2025,
akins+2024,casey+2024,casey+2025,williams+2024,setton+2025}, even though there are
exceptions, with clear detections at MIR wavelengths \citep[e.g.,][]{juodzbalis+2024b,
barro+2025} or even in the FIR \citep[e.g.,][]{barro+2025}.
\target is not X-ray weak (Fig.~\ref{f.xrays}), unlike most \jwst-discovered AGN
at $z>2$ \citep{yue+2024,maiolino+2025x}.

\begin{figure*}[!t]
  \centering
      \includegraphics[width=\textwidth]{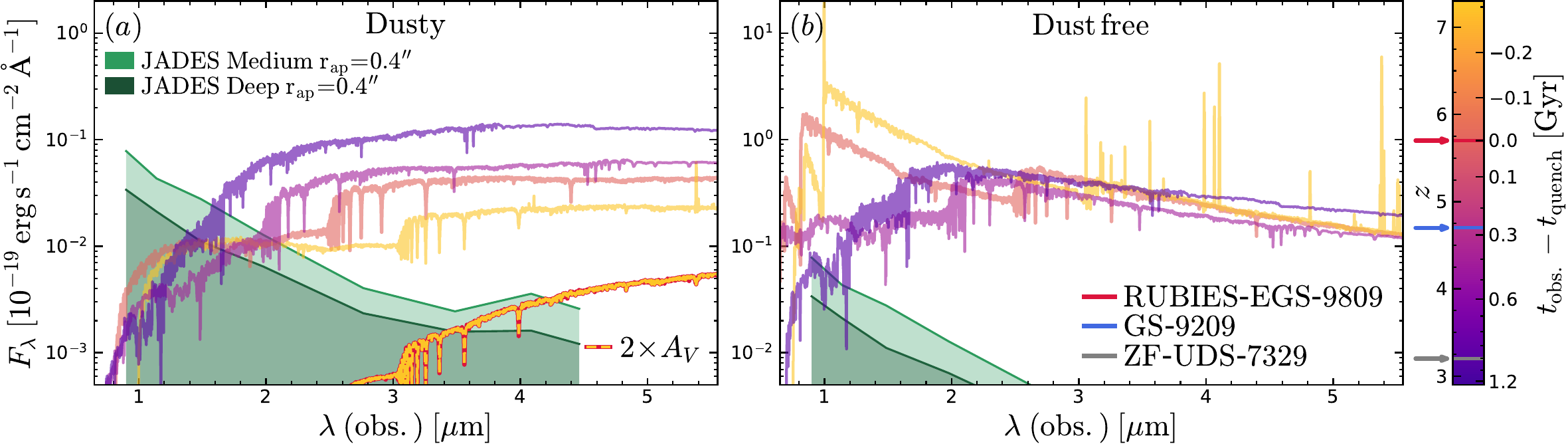}
  {\phantomsubcaption\label{f.future.a}
   \phantomsubcaption\label{f.future.b}}
      \caption{a. Toy model showing the spectral evolution of a dusty
  post-starburst galaxy, between $z=7$ and $z=3$. The reference model reproduces
  the best-fit \cigale model at $z=5.7$. b. Same as panel~\subref{f.future.a}, but
  without dust attenuation. The colorbar is the same for both panels. The lines
  across the colorbar mark the redshifts of \target (red), a massive post-starburst
  and dust-free galaxy \citep[GS-9209;][]{carnall+2023}, and a very old 4000-\AA
  break galaxy \citep[ZF-UDS-7329;][]{glazebrook+2024}. Our proposed model would
  see the galaxy quenching first ($z\gtrsim 5.7$), then removing the dust
  ($z\sim4.5$), thus skipping the UV-bright and quenched phase, which is readily
  detectable, but has not been observed.
  The green shaded regions
  show the 5-\textsigma sensitivities of JADES medium and deep. The dust-obscured,
  star-forming progenitor of \target should be readily detectable in JADES (yellow
  spectrum in panel~\subref{f.future.a}); the lack of reported detections may
  require even higher continuum attenuation than the \cigale value of $A_V=1.8$~mag
  (yellow-red curves).
  }\label{f.future}
\end{figure*}

Interestingly, \target also displays outflows, both in the ionized phase
(Section~\ref{s.r.ss.oution}) and, possibly, in the neutral phase too
(Section~\ref{s.r.ss.outnai}). This is in stark contrast to most LRDs,
whose ionized outflows -- when detected at all -- are weak and more
reminiscent of the modest galaxy-scale winds seen in low-mass star-forming galaxies at high redshift \citetext{cf. \citealp{juodzbalis+2024b,deugenio+2025e} for LRDs and \citealp{xu+2023,carniani+2024a} for galaxies}.
Instead, even considering all the associated systematic uncertainties, the velocity and mass outflow rate of the outflows in \target match what is seen in more massive AGN hosts at lower redshift \citep[e.g.,][]{bertola+2025}.
In the context of all peculiar properties of \target relative
to LRDs, the presence of outflows detected in metal lines is unlikely to be a mere
coincidence. Logically, metal-loaded outflows may be one way to clear paths through
the dense absorbing gas, enabling ionizing radiation to escape, while also enhancing
AGN feedback by increasing the coupling between radiation and dust or metals
\citep[e.g.,][]{fabian+2008}.

Overall, these considerations give confidence to our measurements of an evolved stellar
population and hence a genuinely large value of \mstar. However, we also acknowledge that
higher-resolution, high-quality spectroscopy capable of spectrally resolving the
narrow-line emission is necessary to fully rule out the LRD interpretation.

While in the next sections we discuss the fiducial interpretation that \target is
a massive galaxy, there remains a possibility that we are also witnessing
a transitional AGN between the LRD phase and a `normal' AGN, as recently suggested for similar sources by \citet{fu+2025}.
Assuming that \target was an LRD, detection in X-rays would imply that the intrinsic
SED of LRDs is indeed hard -- e.g. via the presence of a hot corona \citep{hotcor}. This would require a high flux of ionizing radiation,
thus explaining the detection of high ionization lines in some LRDs
\citep{tripodi+2024,tang+2025}. The general lack of X-ray detection would then be in agreement with the high
covering factor proposed by \citet[e.g.,][]{maiolino+2025x}, without requiring an intrinsically soft X-ray spectrum, as posited by some models of super-Eddington accretion \citep{pacucci+2024,madau+haardt2025,lambrides+2024}.

The mix of ionized and neutral metals would be possible if the ISM
was initially starved of ionizing photons \citep[such would be possible
for LRDs;][]{ji+2025a,setton+2025}, but the system is observed right during
the blow-out phase when feedback is starting to clear some of the obscuring
material.

\subsection{A `dusty path' to quiescence}\label{s.disc.ss.dustypath}

In Fig.~\ref{f.future} we show how \target would evolve if it stopped forming stars
at the epoch of observation (marked as $t_\mathrm{obs.}-t_\mathrm{quench}=0$ in the
colorbar). We take as reference the best-fit \cigale model, with $A_V=1.8$~mag, and
evolve it backwards and forwards in time. Panels~\subref{f.future.a}
and~\subref{f.future.b} show dust-attenuated and dust-free spectra, respectively.
The four coloured spectra in
each panel are four snapshots in time; the snapshot at
$t_\mathrm{obs.}-t_\mathrm{quench}=0.3$~Gyr is chosen to be close in time to the
post-starburst galaxy GS-9209 \citep{carnall+2023}, while the
  $t_\mathrm{obs.}-t_\mathrm{quench}=1$~Gyr is near the very old galaxy ZF-UDS-7329
\citep{glazebrook+2024}.
The highest-redshift snapshot (yellow) is the star-forming progenitor (with $2\times$
lower \mstar). In principle, the evolutionary tracks can jump from panel~\subref{f.future.b}
to panel~\subref{f.future.a} by forming dust, and vice versa, dust removal would move the
evolution from panel~\subref{f.future.a} back to \subref{f.future.b}.

Since not all these phases have been observed, we can draw informed hypotheses
on plausible evolutionary paths. As a reference, we show the 5-\textsigma
sensitivity of JADES for both the medium tier ($\sim150~\mathrm{arcmin}^2$) and
deep tier \citep[$\sim32~\mathrm{arcmin}^2$;][]{eisenstein+2023}, calculated
inside a circular aperture with 0.4-arcsec radius \citep{sun+2025}.
Since all four models lie clearly above the JADES detection limits in both panels,
we should in principle be able to detect all these evolutionary stages. Instead,
the star-forming progenitors are missing, in both the dust-obscured and (even
more so) in the dust-free scenarios. A possible explanation is that the dusty progenitors are even more obscured than inferred from \cigale (red-and-yellow spectrum in panel~\subref{f.future.a}).
This interpretation is consistent with the recent discovery of extremely dusty galaxies at
$z\gtrsim8$ \citep{bakx+2025,zavala+2025}, which likely sample the most obscured tail of the progenitor population and may be undercounted in NIRCam-selected spectroscopic follow-up
\citep{sun+2025}.

The bolometric luminosity of this galaxy (incl. the dominant AGN) is of order $10^{13}~\Lsun$,
meaning this galaxy is on par with hot, dust-obscured galaxies discovered by \textit{WISE}
and \textit{Spitzer} \citep[e.g.,][]{eisenhardt+2012,wu+2012}.
\begin{figure*}[!t]
    \centering
	\includegraphics[width=\textwidth]{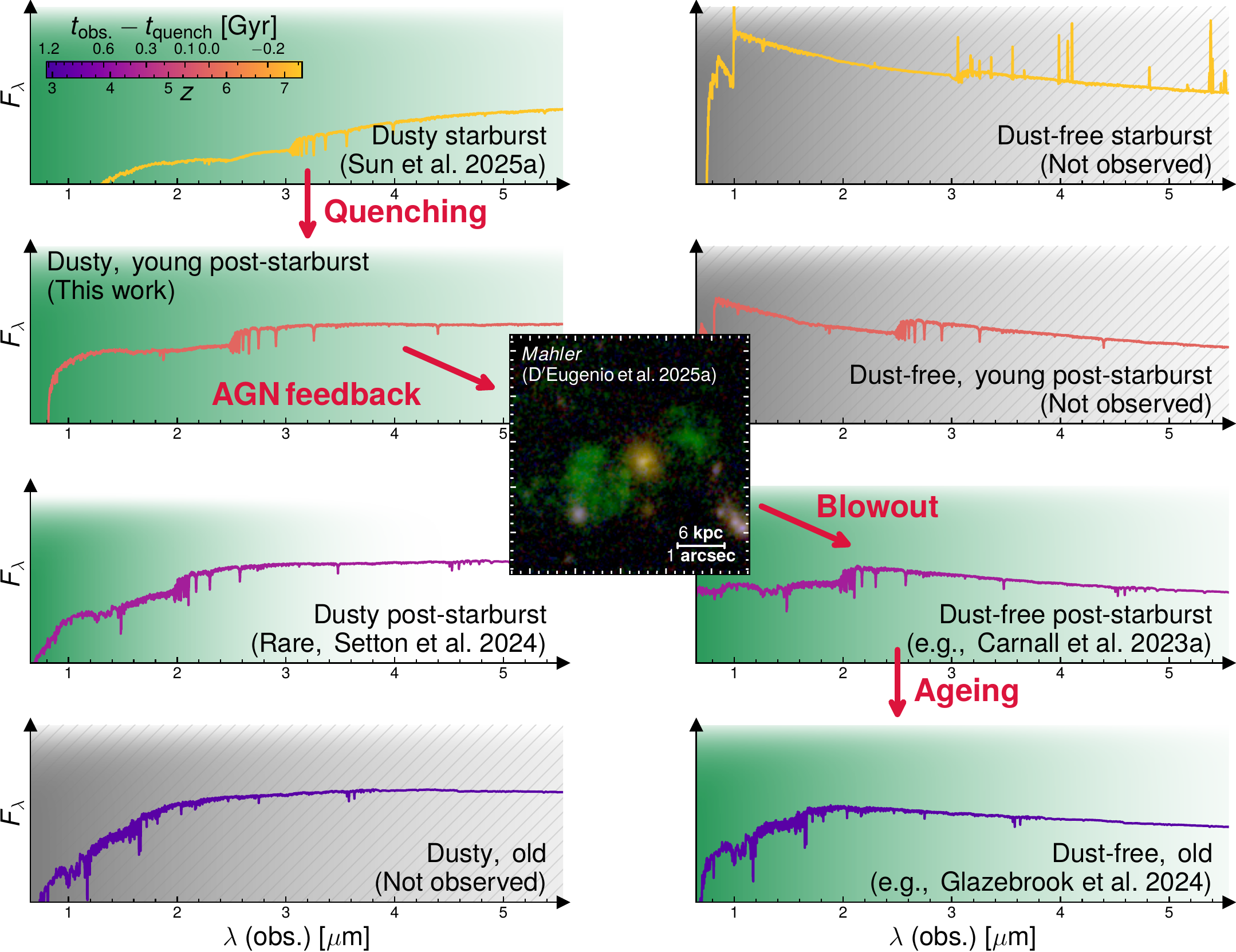}
	\caption{Cartoon of the dusty path to quiescence of massive, early galaxies,
    where we illustrate in shaded green plausible paths, and in grey evolutionary
    phases that are disfavoured by observations.
    The star-forming phase must be dust obscured, since no suitable progenitors
    have been identified, despite their intrinsic brightness (Fig.~\ref{f.future.b}).
    Similarly, young post-starburst systems with ages $\lesssim 50$~Myr must be dust obscured,
    since they have not been observed, even though equal-age galaxies can be detected
    down to 2-dex lower \mstar \citep[e.g.,][]{looser+2024,kuruvanthodi+2024,baker+2025b}.
    Quenching thus happens still in the dust-obscured phase (albeit the dust attenuation
    may be lower than during the starburst phase). When, eventually, dust is removed
    (for instance, by violent AGN feedback, \citealp{deugenio+2025a,perez-gonzalez+2025}),
    the system is revealed to be already in a post-starburst phase, which is commonly
    observed \citetext{e.g., \citealp{carnall+2023,degraaff+2025b}; in some rarer cases, dust
    can linger on during the post-starburst phase \citealp{setton+2024}}. Passive evolution then
    leads to dust-poor (or dust-free) old systems \citep{glazebrook+2024,mcconachie+2025}.
    }\label{f.cartoon}
\end{figure*}

We summarize these considerations in the cartoon of Fig.~\ref{f.cartoon}.
Here we show the two evolutionary sequences side by side (dusty and dust-free),
and colour-code in grey evolutionary phases that are disfavoured by observations.
Green-background phases instead have been directly observed, thus representing a
plausible path. The star-forming progenitors should be heavily obscured starbursts,
with SFR of $\sim100~\Msun~\peryr$, sufficiently low to be missed by shallow ALMA
surveys \citep{sun+2025}. As quenching starts it reduces but --
crucially -- does not fully eliminate dust. This dust-obscured
quenching phase is seen in \target here, and in other systems such as \textit{Hyde}
\citep{schreiber+2018,perez-gonzalez+2025}. AGN feedback is crucial in this phase:
since the SFR is dwindling, it seems unlikely to us that it may drive dust-clearing
outflows at this evolutionary stage, when it was unable to do so earlier, at the
peak of the starburst.
Instead, AGN are known to drive fast outflows in both quenching
\citep{deugenio+2024a,belli+2024} and quiescent galaxies \citep{sun+2025b}, and
are seen \textit{acting} directly in both \target and \textit{Hyde}. The case of
\textit{Mahler}, a $10^{10}$-\Msun galaxy at $z=5.9$ may represent the
`blowout' phase when the AGN just removed a substantial fraction of the ISM.
However, at this stage, the galaxy already has a well developed Balmer break, while
dust can explain the lack of UV-bright Balmer-break galaxies in the mass range
$\mstar>10^{10}~\Msun$. Indeed, young and post-starburst galaxies all
display evidence for fast outflows or even dust removal \citetext{e.g.,
GS-9209, \citealp{ji+2025b}; GS-10578, \citealp{deugenio+2024a}; RUBIES-UDS-Qz7, \citealp{valentino+2025}}.

We can thus propose a `dusty-path' towards quiescence, where dust removal, being
delayed relative to quenching, conceals the UV-bright, quenched phase that is instead
commonly observed in galaxies with 2-dex lower \mstar
\citep{looser+2024,kuruvanthodi+2024,baker+2025b}.
The prevalence of this evolutionary path for the general population of massive
quiescent galaxies at $z\gtrsim3$ is unclear. The lack of massive, UV-bright and
quenched galaxies suggests it should be widespread, but a complete census of dusty,
post-starburst systems is yet to be done, because existing spectroscopic projects
\citep{barrufet+2024,cooper+2025b} selected targets from medium-depth photometry, which may
be too shallow \citep{sun+2025}, even though a few photometric candidates are known
\citep{xiao+2023}.

More generally, the population of heavily obscured, compact galaxies at $z>2$
may be currently under-characterized: on one hand, the bulk of these galaxies may escape
both medium-depth NIRCam and MIRI surveys, and shallow ALMA surveys \citep{sun+2025}. On
the other hand, their extremely red SEDs make it generally difficult to infer the redshift
from photometry alone, let alone infer the full diversity of star-formation histories
and AGN properties \citep{cooper+2025b}.
Deep NIRSpec observations to follow up faint, dust-obscured candidates are a compelling path forward.

\section{Summary and Conclusions}\label{s.conc}

In this work, we present a full spectro-photometric analysis of the dust-reddened source \target, originally identified as an X-ray emitter (Section~\ref{s.data}), and spectroscopically confirmed with \jwst/NIRSpec by RUBIES.
\begin{itemize}
  \item Morphologically, the source is compact yet clearly resolved
  in all NIRCam bands, including both continuum- and line-dominated bands (Section~\ref{s.i.ss.morph}). We find rest-frame optical half-light radii of order $\re \sim 0.4\text{--}0.6$~kpc, consistent with massive, quiescent galaxies at comparable redshifts.
  \item Modelling the SED with \cigale, we find clear evidence of a Balmer break in the photometry
  (Figs.~\ref{f.sizes} and~\ref{f.sizewave}), in agreement with the NIRSpec data, where we detect the
  break with 4-\textsigma significance (Fig.~\ref{f.bbreak}).
  \item The AGN is also clearly seen at MIR wavelengths and in the radio. Cold dust emission is tentatively detected by JCMT/SCUBA-2 at 850~\mum. \cigale requires a dust-obscured
 AGN, and interprets the Balmer break as evidence for an evolved stellar population,
  with $\mstar = 10^{10.5\text{--}11}~\Msun$ (Section~\ref{s.i.ss.sed}).
  \item The prism spectrum displays bright and broad emission, which we identify as broad \Halpha from an AGN broad-line region. We model the line as the sum of two Gaussians, and we infer a combined $FWHM = 1500_{-100}^{+200}~\kms$, while the broadest component has $FWHM = 3800_{-300}^{+400}~\kms$.
  Due to the lack of medium-resolution coverage, we cannot readily deblend \Halpha from \NIIall.
  \item The grating covers the \Hbeta--\OIIIall spectral range, where we detect broad \OIII, which we interpret as an ionized outflow. The resulting
  mass outflow rate is $\log(\Mdotout[{\OIII}]/(\Msun\,\peryr)) = 0.36\pm0.04$, with much larger systematic uncertainties (Section~\ref{s.r.ss.oution}).
  \item We detect deep \NaIall absorption, tentatively blueshifted (2~\textsigma), with high $\rm{EW}(\NaIall)=17\pm2~\AA$ -- a value so large as to require an ISM origin. If interpreted as an outflow, this yields a neutral-phase mass outflow rate $\log(\Mdotout[{\NaI}]/(\Msun\,\peryr)) = 2.3^{+0.2}_{-0.3}$.
  \item The absorption-corrected X-ray luminosity is in line with scaling relations typical of broad-line AGN (Fig.~\ref{f.xrays}), unlike X-ray weak LRDs.
\end{itemize}

We interpret this source as a massive, evolved galaxy and a strong quenching candidate. The causes underlying the decreasing SFR trend cannot be untangled with our data, and span
all plausible ranges, from AGN or star-formation driven feedback, to gas consumption. The emission-line model grid favours high intrinsic BLR Balmer decrements, but leaves the recent SFR too uncertain to establish quiescence on its own. The current outflows observed in the source are consistent with an AGN-driven outburst, given their coincidence with a high-luminosity AGN, but confirmation would require further observations.
We argue that \target may be caught in a transition phase, similar to other such systems observed near or during the removal of substantial amounts of ISM \citep{parlanti+2024,deugenio+2025c,perez-gonzalez+2025}, some of which are confirmed to host evolved stellar populations \citep{deugenio+2025c,perez-gonzalez+2025} and all of which are reported to host luminous, dust-obscured AGN.
This population represents one step closer to connecting massive, quiescent galaxies at $z=3\text{--}4$ to their dust-obscured, star-forming progenitors \citep{sun+2025}. If this quenched and dust-obscured phase was common, it could explain the lack of dust-free, intermediate-age post-starburst galaxies between the NIRCam-dark star-forming progenitors and their dust-free, quiescent descendants.
An accurate characterization of their SED and SFH is crucial for understanding if and how AGN feedback is able to quench star formation in massive, early galaxies.

\section*{Data availability}

This work uses publicly available data: NIRCam imaging is from CEERS, using the
reduction from the DAWN JWST Archive (DJA).
NIRSpec spectroscopy is from the RUBIES and CAPERS programmes, using the DJA v4
reduction. The non-\jwst data are taken from AEGIS-XD (\textit{Chandra}),
\textit{Spitzer} PIDs 8, 41023 and 61042 (\textit{Spitzer}/IRAC), FIDEL
(\textit{Spitzer}/MIPS), JCMT/SCUBA-2, and AEGIS20 (VLA).

\section*{Acknowledgments}

The authors thank Max Pettini, Benjamin D. Johnson, Giovanni Mazzolari, and Elena and Roberto Terlevich for helpful discussions.
FDE, MB, RM, IJ, XJ, RGP and JS acknowledge support by the Science and Technology Facilities Council (STFC), by the ERC through Advanced Grant 695671 ``QUENCH'', and by the UKRI Frontier Research grant RISEandFALL.
RM also acknowledges funding from a research professorship from the Royal Society.
EB acknowledges financial support from the MIRACLE INAF 2024 GO grant ``A JWST/MIRI MIRACLE: Mid-IR Activity of Circumnuclear Line Emission'' and the Ricerca Fondamentale INAF 2024 under the project 1.05.24.07.01 MiniGrant RSN1.
IJ also acknowledges support by the Huo Family Foundation through a P.C. Ho PhD Studentship.
SC and EP acknowledge support by European Union's HE ERC Starting Grant No. 101040227 - WINGS.
This work is sponsored by the National Key R\&D Program of China under grant No. 2022YFA1605300 and the National Natural Science Foundation of China (NSFC) grant Nos. 12273051 and 11933003. Support for this work is also partly provided by the CASSACA. YL gratefully acknowledges support from the Royal Society International Exchanges Scheme (IES\textbackslash R1\textbackslash 211140) and the Chinese Academy of Sciences President's International Fellowship Initiative (grant No. 2022VMB0004).
MP acknowledges support from the Kavli Institute for Cosmology, Cambridge and the Kavli Foundation.
MB acknowledges support from the Next Generation EU funds within the National Recovery and Resilience Plan (PNRR), Mission 4 - Education and Research, Component 2 - From Research to Business (M4C2), Investment Line 3.1 - Strengthening and creation of Research Infrastructures, Project IR0000034 – ``STILES - Strengthening the Italian Leadership in ELT and SKA''.

\bibliographystyle{aasjournal}
\bibliography{angry}

\begin{appendix}

\section{Possible Metal absorption}\label{a.metal}

Our continuum model uses an additive step function to capture the observed
continuum depression highlighted by the green vertical band in Fig.~\ref{f.data.b}
and~\subref{f.data.c}. In this appendix we summarize the measurement, clarify why
we favour an absorption-based interpretation, and conclude by providing an
empirical comparison to other sources.

The prism spectrum shows a sharp drop in flux at rest-frame wavelengths
shortward of $\lambda \approx 6250~\AA$ (green vertical band in
Fig.~\ref{f.data.b}). The same feature is also tentatively seen in the G395M
spectrum (Fig.~\ref{f.data.c}), a fact that disfavours blended, unresolved
emission lines (such as \OIall and \SIIIL[6312]), and rules out an artefact,
such as due to the data-reduction pipeline or to spoilers and failed-open or
disobedient shutters \citep{ferruit+2022}, spectral overlaps \citep[e.g.,][]{
degraaff+2025a,deugenio+2025f}, and short circuits \citep[e.g.,][]{deugenio+2025b}.

Denoting the 2\textsuperscript{nd}-order polynomial representing the background as
$\mathcal{P}_2$ (Section~\ref{s.anspec}), the additive step function
$\mathcal{H}$ is given by
\begin{equation}\label{eq.hump}
  \mathcal{H}(\lambda) \equiv
  \begin{cases}
      \beta_\mathrm{h} (\lambda - \lambda_\mathrm{h,0}) \\
      \hspace{1cm} + (A_\mathrm{h} - 1)\cdot \mathcal{P}_2(\lambda) & \lambda_\mathrm{h,0} \leq \lambda < \lambda_\mathrm{h,1}\\
      0 & \text{otherwise}.
   \end{cases}
\end{equation}
This way, $A_\mathrm{h}$\blfootnote{\textcolor{white}{Where, following Wilder \& Brooks (1974), `h' stands for `hump' -- What hump?}}
represents the amplitude of the discontinuity at $\lambda = \lambda_\mathrm{h,0}$,
with $A_\mathrm{h} = 1$ corresponding to no discontinuity. All of $A_\mathrm{h}$, $\beta_\mathrm{h}$, $\lambda_\mathrm{h,0}$ and $\lambda_\mathrm{h,1}$ are free parameters. Depending on their values, there can be a
second discontinuity at $\lambda = \lambda_\mathrm{h,1}$, with amplitude equal to
$\beta_\mathrm{h} (\lambda_\mathrm{h,1}-\lambda_\mathrm{h,0}) + (A_\mathrm{h}-1) \mathcal{P}_2(\lambda_\mathrm{h,1})$.
The above definition focuses on the region near $\lambda_\mathrm{h,0}$, where we
decouple the presence or absence of the discontinuity from the local value of the
background.

To quantify its statistical significance, we fit this spectral region
with the model of Section~\ref{s.anspec}. The probability that no drop in flux
is present is $P(A_\mathrm{h}\leq 1)<0.001$ (3-\textsigma significance), and
the model without this step function performs considerably worse, with
$\Delta \text{BIC}=74$ (Table~\ref{t.modelgrid}).
In the prism spectrum, the corresponding depression removes of order 15~percent
of the local continuum flux across the interval
$\lambda \simeq 6160\text{--}6250$~\AA.

\begin{center}
    \begin{figure}[!t]
      {\phantomsubcaption\label{f.capershump.a}
       \phantomsubcaption\label{f.capershump.b}
       \phantomsubcaption\label{f.capershump.c}
       \phantomsubcaption\label{f.capershump.d}
       \phantomsubcaption\label{f.capershump.e}
       \phantomsubcaption\label{f.capershump.f}
       \phantomsubcaption\label{f.capershump.g}
       \phantomsubcaption\label{f.capershump.h}}
        \centering
    	\includegraphics[width=\columnwidth]{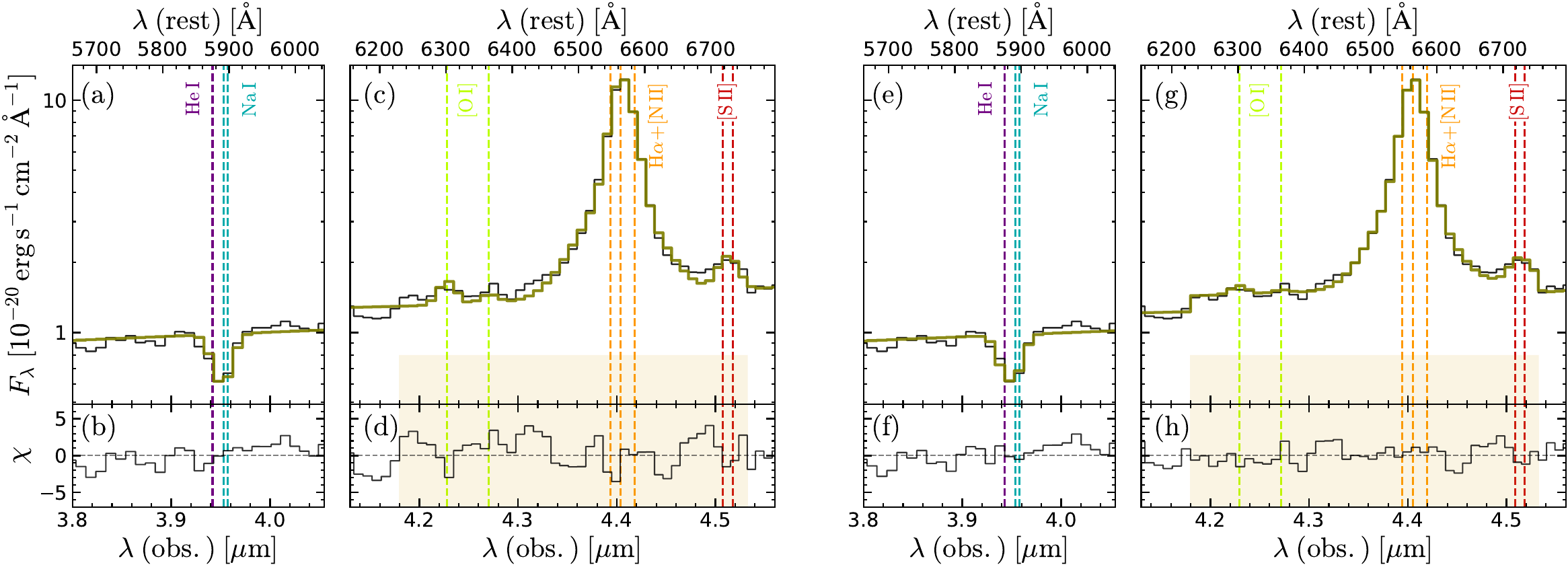}
    	\caption{The flux drop near $\lambda=6250~\AA$ is weaker but still
        present in the higher-SNR CAPERS spectrum.
        Panels~\subref{f.capershump.a}--\subref{f.capershump.d} show the
        best-fit model without the $\mathcal{H}$ term, while
        panels~\subref{f.capershump.e}--\subref{f.capershump.h} show the
        model that includes $\mathcal{H}$. The lines and colours are the same
        as Fig.~\ref{f.fit}; the shaded region shows the support of the
        $\mathcal{H}$ function, highlighting the strong residuals in
        panels~\subref{f.capershump.d} vs~\subref{f.capershump.h}.
        }\label{f.capershump}
    \end{figure}
\end{center}

Albeit weaker, the feature seems also present in the CAPERS spectrum
(Fig.~\ref{f.data.b}), which suggests a spatial association with the RUBIES
shutter position -- perhaps a preference for the central regions of the source.
When fitting the CAPERS spectrum with the same setup as Section~\ref{s.anspec}, we confirm the
preference for the fit that includes the $\mathcal{H}$ model, with $\Delta\,\BIC
= 111$, due to the higher SNR of the CAPERS data (Fig.~\ref{f.capershump}).

If confirmed by deeper and higher-resolution observations, this feature may
be due to low-ionization metal absorption in the continuum, e.g. due to \CaI
and \FeI.

As an alternative to the fiducial model of Eq.~\ref{eq.hump}, we also
attempted to explain this feature with permitted \FeII emission, but this
model is disfavoured by the data (Appendix~\ref{a.feii}), which reinforces
the hypothesis that the feature represents true absorption, rather than
(local) absence of spectrally unresolved emission lines.
Of course, this conclusion remains tentative, but it is supported by the
very strong \NaIall absorption (Section~\ref{s.r.ss.outnai}), which
independently points to a metal-rich, low-ionization ISM along the line of sight.

\begin{center}
    \begin{figure}[!t]
      {\phantomsubcaption\label{f.metals.a}
       \phantomsubcaption\label{f.metals.b}}
        \centering
    	\includegraphics[width=\columnwidth]{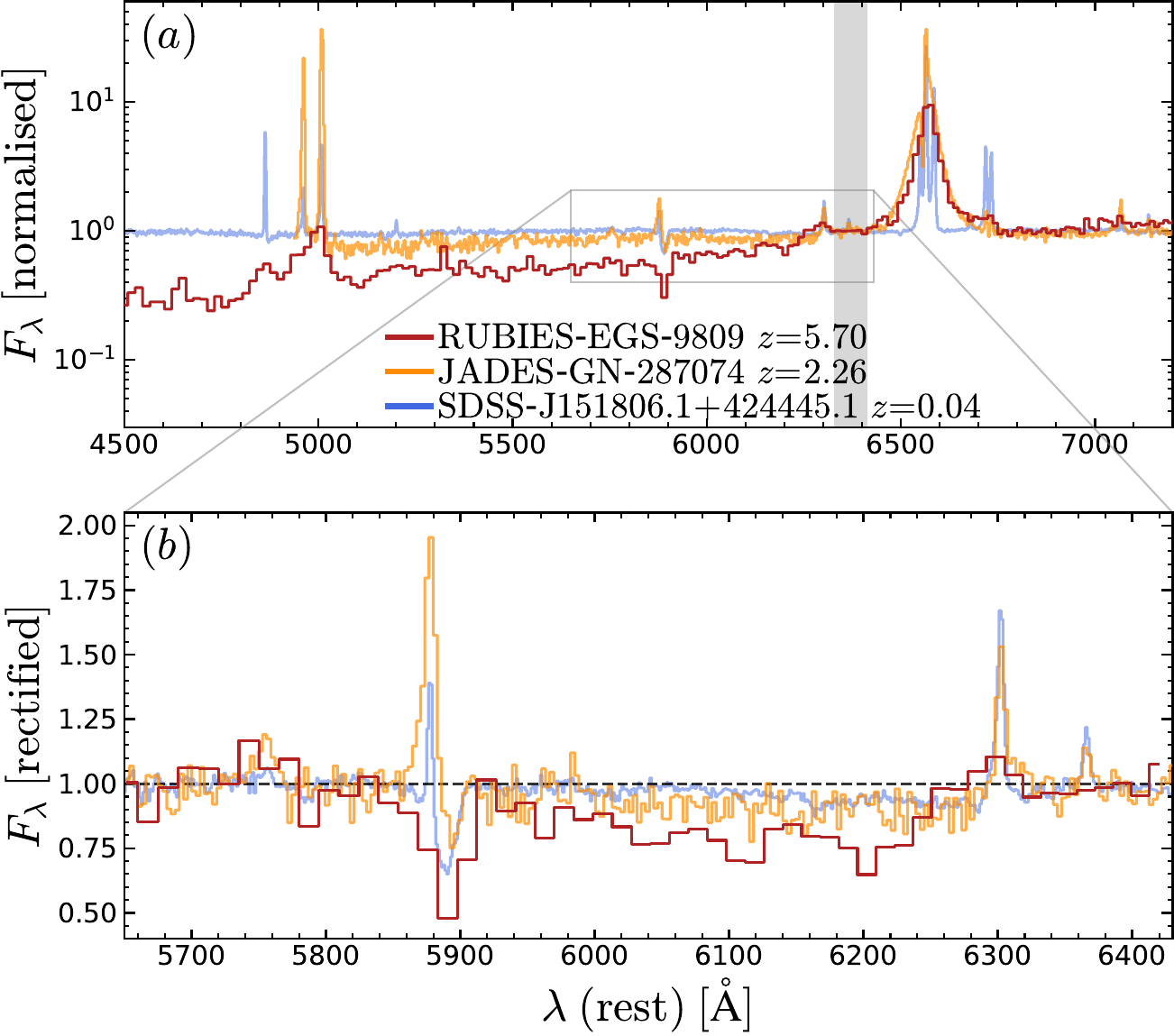}
    	\caption{The flux drop at $\lambda \sim 6250~\AA$ we report in \target
        is also seen in other sources associated with LRDs and/or outflows.
        Panel~\subref{f.metals.a} shows two comparison objects, \rosetta
        \citep[a luminous LRD;][]{juodzbalis+2024b}, and \sdss,
        a dusty starburst with a neutral-gas outflow traced by \NaIall. The
        spectra have been normalized using the median flux in the vertical
        grey interval.
        Panel~\subref{f.metals.b} displays a detail of the absorption, enhanced
        by dividing by a continuum model. Note that \target has both the strongest
        absorption feature and the highest-EW \NaIall absorption.
        }\label{f.metals}
    \end{figure}
\end{center}

The absorption feature near $\lambda=6250~\AA$ is also seen in other
galaxies. In Fig.~\ref{f.metals}, we compare \target to two other galaxies:
\sdss at $z=0.04$ and \rosetta at $z=2.26$ \citep{juodzbalis+2024b}. Both
comparison sources display the same spectral drop, albeit with considerably
smaller amplitude. In addition, both show clearly detected \NaI absorption,
while \rosetta is a well-known LRD AGN \citep{juodzbalis+2024b}. Motivated by
this similarity, we define a simple phenomenological measure of the continuum
depression, the Index for Metal-absorbing BLR-AGN (hereafter, IMB), following
the spirit of the Lick indices \citep{worthey+1994}. The IMB is intended only
as an empirical comparison tool between spectra, rather than an actual metal-line
index. We define a blue continuum as the median of the source continuum within
the wavelength range $5600\text{--}5800$~\AA, and a red continuum as the median within
the wavelength range $6317\text{--}6440$~\AA, masking the spectral region around
$\sim 6360$\AA\ (specifically: $6357\text{--}6375$ \AA) to avoid the \OI line
contamination in the continuum estimate.
We assume the interval midpoints as reference wavelengths for fitting a linear
relation approximating the continuum around the absorption feature at $\lambda
\sim 6250$ \AA. We exploit such relation to retrieve the unabsorbed continuum
estimate at reference wavelength $6118$~\AA, which we compare with the measured
absorbed flux evaluated as the median value within the absorption interval
$6160\text{--}6250$~\AA.
From the prism spectrum of \target, we retrieve an absorption strength of
$1.18 \pm 0.42$, meaning that $\sim 15$~percent of the $\lambda \sim 6250$
\AA flux is absorbed.
Following the same procedure, we also analyse \sdss and \rosetta, and find
absorption strengths of $1.040 \pm 0.027$ and $1.12 \pm 0.18$, respectively.

We then search for possible correlations with the nearby \NaI absorption
line strength. With this method, we measure rest-frame \NaI equivalent widths of
$13 \pm 3$~\AA, $5.28 \pm 0.05$~\AA and $2.9 \pm 0.4$~\AA for
\target, \rosetta, and \sdss, respectively.
Thus, within this small comparison set, \target shows both the strongest
continuum depression and the strongest \NaI absorption. This is only suggestive,
given the small, heterogeneous sample, but it is consistent
with the idea that both features trace unusually strong columns of cool,
metal-rich gas. A secure physical interpretation will require deeper
medium-resolution spectroscopy covering this wavelength range in a larger
sample of dusty AGN and \NaI absorbers.

\section{\texorpdfstring{\FeII}{FeII} emission tests}\label{a.feii}

To test if the continuum drop at $6160\text{--}6250$~\AA can be reproduced
by broad \FeII emission instead of absorption, we re-run the fiducial spectral
model by adding \FeII templates \citep{veron+cetty2004}. The \FeII contribution
is described by three free parameters: the template amplitude $A_{\FeII}$, the
attenuation applied to the \FeII component, $A_{V,\FeII}$ \citetext{assuming the
\citetalias{gordon+2003} law, as before}, and the iron-line width, $FWHM_{\FeII}$.

This model is disfavoured by the data ($\Delta\,\BIC = 41$; Table~\ref{t.modelgrid})
for two separate reasons. First, the wavelength distribution of the strongest
\FeII lines does not match the features observed near 6200~\AA at all.
Moreover, the model cannot simultaneously reproduce a relatively sharp drop
(Fig.~\ref{f.data.b}) and the lack of resolved nearby emission features: adequately
fitting the drop would require low $FWHM_{\FeII}$, but the smooth observed spectrum
needs $FWHM_{\FeII}\gtrsim1000~\kms$.

The fit therefore converges to a broad component,
$FWHM_{\FeII}=3400^{+900}_{-800}~\kms$, which successfully smooths out any
resolved peaks, but at the same time still cannot reproduce the observed flux
drop at $\lambda\sim6250$~\AA\ (Fig.~\ref{f.data.b}). The fitted template
amplitude is marginally non-zero, $A_{\FeII}=28^{+11}_{-10}$, indicating that
this model can indeed explain a small amount of residual structure, but the
poor fit quality means that \FeII alone cannot fully explain the feature.
Finally, the fit requires substantial attenuation of the \FeII component,
$A_{V,\FeII}=4.3^{+0.4}_{-0.4}$~mag, which further suppresses any associated
bluer \FeII emission. We therefore disfavour the \FeII interpretation and favour
an absorption-based model (Appendix~\ref{a.metal}).

\end{appendix}

\end{document}

%% file: config/fde_emlines.tex
\usepackage{textgreek}
\usepackage{xargs}
\usepackage{xspace}

\newcommand{\Halpha}{\text{H\,\textalpha}\xspace}
\newcommand{\Hbeta}{\text{H\,\textbeta}\xspace}
\newcommand{\Hgamma}{\text{H\,\textgamma}\xspace}

\newcommandx{\permittedEL}[6][1=O,2=III,3=,4=,5=,6=]{\text{{#1}\,{\sc {#2}}{#3}{#4}{#5}{#6}}\xspace}
\newcommandx{\semiforbiddenEL}[6][1=O,2=III,3=,4=,5=,6=]{\text{{#1}\,{\sc{#2}}]{#3}{#4}{#5}{#6}}\xspace}
\newcommandx{\forbiddenEL}[6][1=O,2=III,3=,4=,5=,6=]{\text{[{#1}\,{\sc{#2}}]{#3}{#4}{#5}{#6}}\xspace}

\newcommandx{\NVL}[1][1=1243]{\permittedEL[N][v][\textlambda][#1]}
\newcommandx{\NVall}{\permittedEL[N][v][\textlambda][\textlambda][1239,][1243]}

\newcommandx{\CIIL}[1][1=232x]{\semiforbiddenEL[C][ii][\textlambda][#1]}
\newcommandx{\CIIall}{\semiforbiddenEL[C][ii][\textlambda][\textlambda][2324--][2329]}

\newcommandx{\NIVL}[1][1=1486]{\semiforbiddenEL[N][iv][\textlambda][#1]}

\newcommandx{\CIVL}[1][1=1550]{\permittedEL[C][iv][\textlambda][#1]}

\newcommandx{\HeIIL}[1][1=1640]{\permittedEL[He][ii][\textlambda][#1]}

\newcommandx{\semiOIIIL}[1][1=1666]{\semiforbiddenEL[O][iii][\textlambda][#1]}

\newcommandx{\NIIIL}[1][1=1750]{\semiforbiddenEL[N][iii][\textlambda][#1]}

\newcommandx{\CIII}{\semiforbiddenEL[C][iii]}
\newcommandx{\CIIIL}[1][1=1909]{\semiforbiddenEL[C][iii][\textlambda][#1]}

\newcommandx{\NeIVL}[1][1=2424]{\forbiddenEL[Ne][iv][\textlambda][#1]}

\newcommandx{\MgIIL}[1][1=2803]{\permittedEL[Mg][ii][\textlambda][#1]}

\newcommandx{\NeVL}[1][1=3426]{\forbiddenEL[Ne][v][\textlambda][#1]}

\newcommandx{\OIIL}[1][1=3726]{\forbiddenEL[O][ii][\textlambda][#1]}
\newcommand{\OIIall}{\forbiddenEL[O][ii][\textlambda][\textlambda][3726,][3729]}

\newcommandx{\NeIIIL}[1][1=3869]{\forbiddenEL[Ne][iii][\textlambda][#1]}

\newcommand{\OIII}{\forbiddenEL[O][iii]}
\newcommandx{\OIIIL}[1][1=5007]{\forbiddenEL[O][iii][\textlambda][#1]}
\newcommand{\OIIIall}{\forbiddenEL[O][iii][\textlambda][\textlambda][4959,][5007]}

\newcommandx{\NIL}[1][1=5200]{\forbiddenEL[N][i][\textlambda][#1]}

\newcommand{\OI}{\forbiddenEL[O][i]}
\newcommandx{\OIL}[1][1=6300]{\forbiddenEL[O][i][\textlambda][#1]}
\newcommand{\OIall}{\forbiddenEL[O][i][\textlambda][\textlambda][6300,][6364]}

\newcommandx{\HeIL}[1][1=5875]{\permittedEL[He][i][\textlambda][#1]}

\newcommandx{\OIresL}[1][1=8446]{\permittedEL[O][i][\textlambda][#1]}

\newcommand{\NII}{\forbiddenEL[N][ii]}
\newcommandx{\NIIL}[1][1=6583]{\forbiddenEL[N][ii][\textlambda][#1]}
\newcommand{\NIIall}{\forbiddenEL[N][ii][\textlambda][\textlambda][6548,][6583]}

\newcommandx{\SIIL}[1][1=6716]{\forbiddenEL[S][ii][\textlambda][#1]}
\newcommand{\SIIall}{\forbiddenEL[S][ii][\textlambda][\textlambda][6716,][6731]}

\newcommandx{\OIIAuL}[1][1=7325]{\forbiddenEL[O][ii][\textlambda][#1]}
\newcommand{\OIIAuall}{\forbiddenEL[O][ii][\textlambda][\textlambda][7319--][7331]}

\newcommandx{\SIIIL}[1][1=9532]{\forbiddenEL[S][iii][\textlambda][#1]}

\newcommand{\FeII}{\permittedEL[Fe][ii]}

\newcommand{\NaI}{\permittedEL[Na][i]}
\newcommand{\CaI}{\permittedEL[Ca][i]}
\newcommand{\FeI}{\permittedEL[Fe][i]}
\newcommandx{\NaIL}[1][1=5890]{\permittedEL[Na][i][\textlambda][#1]}
\newcommand{\NaIall}{\permittedEL[Na][i][\textlambda][\textlambda][5890,][5896]}

\newcommandx{\CIIFIRL}{\forbiddenEL[C][ii][\textlambda][158\,\mum]}


%% file: config/macros.tex
\usepackage{etoolbox}
\makeatletter
\newcommand\sendemail[4]{
\edef\@tempa{mailto:#1?subject=#2&body=#3 }%
\edef\@tempb{\expandafter\html@spaces\@tempa\@empty}%
\href{\@tempb}{#4}}

\catcode\%=11
\def\html@spaces#1 #2{#1
\catcode\%=14
\makeatother

\newcommand{\todo}[1]{\textcolor{magenta}{[#1]}}
\newcommand{\orcid}[2]{\href{http://orcid.org/#2}{#1}}
\newcommand{\orcidsymb}[2]{\href{http://orcid.org/#2}{#1\adjustbox{trim={-.15\width} {0\height} {-.15\width} {0\height},clip}{\includegraphics[height=10pt]{figs/orcid.pdf}}}}
\newcommand{\uat}[2]{#1}

\newcommand{\citationneeded}{\textcolor{ForestGreen}{$^{\rm citation\;needed}$}}
\let\oldtextsigma\textsigma
\renewcommand{\textsigma}{\oldtextsigma\xspace}
\let\oldAA\AA
\renewcommand{\AA}{\text{\oldAA}\xspace}
\let\oldtextdegree\textdegree
\renewcommand{\textdegree}{\oldtextdegree\xspace}

\newcommand{\kms}{\ensuremath{\mathrm{km\,s^{-1}}}\xspace}
\newcommand{\Msun}{\ensuremath{{\rm M}_\odot}\xspace}
\newcommand{\Zsun}{\ensuremath{{\rm Z}_\odot}\xspace}
\newcommand{\Lsun}{\ensuremath{{\rm L}_\odot}\xspace}
\newcommand{\yr}{\ensuremath{{\rm yr}}\xspace}
\newcommand{\Myr}{\ensuremath{{\rm Myr}}\xspace}
\newcommand{\Gyr}{\ensuremath{{\rm Gyr}}\xspace}
\newcommand{\peryr}{\ensuremath{{\rm yr^{-1}}}\xspace}
\newcommand{\mum}{\text{\textmu m}\xspace}
\newcommand{\kpc}{\text{kpc}\xspace}
\newcommand{\ZH}{\text{[Z/H]}\xspace}
\newcommandx{\pcm}[1][1=3]{\ensuremath{\mathrm{cm}^{-#1}}\xspace}	

\newcommandx{\lambdar}[2][1=R,2=]{\ensuremath{\lambda_{\rm {#1}}{#2}}\xspace}
\newcommand{\eps}{\ensuremath{\epsilon}\xspace}
\newcommand{\mstar}{\ensuremath{M_\star}\xspace}
\newcommand{\mdyn}{\ensuremath{M_\mathrm{dyn}}\xspace}
\newcommand{\re}{\ensuremath{R_\mathrm{e}}\xspace}
\newcommand{\vstar}{\ensuremath{v_\star}\xspace}
\newcommand{\vnai}{\ensuremath{v_{\NaI}}\xspace}
\newcommand{\sigmastar}{\ensuremath{\sigma_\star}\xspace}
\newcommand{\sigmaestar}{\ensuremath{\sigma_{\star,\mathrm{e}}}\xspace}
\newcommand{\vperc}[1]{\ensuremath{v_{#1}}\xspace}

\newcommand{\vesc}{\ensuremath{v_\mathrm{esc}}\xspace}
\newcommand{\nelec}{\ensuremath{n_\mathrm{e}}\xspace}
\newcommand{\Telec}{\ensuremath{T_\mathrm{e}}\xspace}
\newcommand{\Rout}{\ensuremath{R_\mathrm{out}}\xspace}
\newcommand{\vout}{\ensuremath{v_\mathrm{out}}\xspace}
\newcommandx{\Mout}[2][1=,2=]{\ensuremath{M_{\mathrm{out}{#2}}^{#1}}\xspace}
\newcommandx{\Mdotout}[2][1=,2=]{\ensuremath{\dot{M}_{\mathrm{out}{#2}}^{#1}}\xspace}

\newcommandx{\fluxdcgs}[2][1=-20,2=\times]{\ensuremath{{#2} 10^{#1}~\mathrm{erg\,s^{-1}\,cm^{-2}\,\AA^{-1}}}\xspace}
\newcommandx{\fluxcgs}[2][1=-20,2=\times]{\ensuremath{{#2}10^{#1}~\mathrm{erg\,s^{-1}\,cm^{-2}}}\xspace}
\newcommandx{\powercgs}[1][1=44]{$\times 10^{#1}$~erg~s$^{-1}$\xspace}
\newcommandx{\ergs}{\ensuremath{\mathrm{erg\,s^{-1}}}\xspace}
\newcommand{\Av}{\ensuremath{A_V}\xspace}

\newcommand{\cigale}{{\sc cigale}\xspace}
\newcommand{\jwst}{\textit{JWST}\xspace}
\newcommand{\hst}{\textit{HST}\xspace}
\newcommand{\ppxf}{{\sc ppxf}\xspace}
\newcommand{\prospector}{{\sc prospector}\xspace}
\newcommand{\emcee}{{\sc emcee}\xspace}
\newcommand{\cloudy}{{\sc cloudy}\xspace}
\newcommand{\pyneb}{{\sc pyneb}\xspace}
\newcommandx{\mappings}[1][1=]{{\sc mappings{#1}}\xspace}
\newcommand{\galfit}{{\sc galfit}\xspace}
\newcommand{\qubespec}{{\sc qubespec}\xspace}
\newcommand{\pysersic}{{\sc pysersic}\xspace}

\newcommand{\blackthunder}{BlackTHUNDER\xspace}
\newcommand{\Mdynvalue}{$\Mdyn = 2.0\pm0.5 \times 10^{11}$~\MSun}

\defcitealias{gordon+2003}{G03}

%% file: table_photometry.tex
\begin{table}
    \setlength{\tabcolsep}{3.5pt}

    \caption{Summary of broad-band photometry for \target.
    }\label{t.phot}
    \begin{tabularx}{\columnwidth}{llXXr}
  \hline
    Instrument & Filter & Wavelength & Flux                             & Ref. \\
               &        &            & [\textmu Jy]                     &      \\
  \hline
    \textit{JWST}/NIRCam  & ---      & 0.9--4.4 \mum  & ---             & $^a$ \\
    \textit{Spitzer}/IRAC & Ch3      & 5.8 \mum       & $3.40 \pm 0.90$ & $^b$ \\
    \textit{Spitzer}/IRAC & Ch4      & 8.0 \mum       & $10.9 \pm 1.1$  & $^b$ \\
    \textit{Spitzer}/MIPS & MIPS24   & 24  \mum       & $140 \pm 14$    & $^c$ \\
    JCMT/SCUBA-2          & 850-\mum & 850 \mum       & $480 \pm 120$   & $^d$ \\
    VLA                   & Band L   & 20 cm          & $99 \pm 13$     & $^e$ \\
  \hline
    \end{tabularx}
\textit{Notes.} $^a$ Table~\ref{t.sizes}, source `9809';
$^b$ \citet{nandra+2015};
$^c$ FIDEL (Far Infrared Deep Extragalactic Legacy Survey);
$^d$ \citet{zavala+2018},
$^e$ \citet{willner+2012}.
\end{table}

%% file: table_sizes_dja_single.tex
\begin{table*}
    \setlength{\tabcolsep}{2.5pt}
    \caption{Morphological parameters of \target and the two nearby sources (all assumed at $z=5.7$), derived from \pysersic modelling. Each group of three rows corresponds
    to a single NIRCam filter, with the rows showing \target, 9809B, and 9809C, respectively.
    The columns report the marginalized posterior distributions on the model parameters, with the median and
    16\textsuperscript{th}--84\textsuperscript{th} percentile range.
    }\label{t.sizes}
    \begin{tabularx}{\textwidth}{lXXXXXXXX}
\hline
Filter              & Source& Type     & Flux               & \re        & \re         &        $n$      &          $b/a$       &        P.A.      \\
                    &       &          & [nJy]             & [arcsec]   & [kpc]       &                 &                      &       [rad]      \\
\hline
\multirow{3}{*}{F090W} & 9809  & S\'ersic & $0.74^{+0.12}_{-0.07}$ & $(<0.33)$ & $(<1.9)$ & $4^{+3}_{-2}$ & $0.4^{+0.3}_{-0.3}$ & $1.4^{+1.1}_{-0.9}$\\
                      & 9809B & S\'ersic & $34^{+4}_{-2}$ & $0.065^{+0.009}_{-0.009}$ & $0.39^{+0.05}_{-0.05}$ & $1.1^{+0.7}_{-0.4}$ & $0.73^{+0.08}_{-0.10}$ & $2.5^{+0.1}_{-0.1}$\\
                      & 9809C & S\'ersic & $33^{+3}_{-3}$ & $0.091^{+0.012}_{-0.008}$ & $0.55^{+0.07}_{-0.05}$ & $0.74^{+0.22}_{-0.06}$ & $0.73^{+0.07}_{-0.10}$ & $2.84^{+0.08}_{-0.09}$\\
\hline
\multirow{3}{*}{F115W} & 9809  & S\'ersic & $12^{+3}_{-2}$ & $0.15^{+0.04}_{-0.03}$ & $0.9^{+0.2}_{-0.2}$ & $1.0^{+0.9}_{-0.2}$ & $0.3^{+0.2}_{-0.2}$ & $1.7^{+0.9}_{-0.9}$\\
                      & 9809B & S\'ersic & $42^{+3}_{-2}$ & $0.074^{+0.005}_{-0.006}$ & $0.44^{+0.03}_{-0.03}$ & $0.71^{+0.09}_{-0.05}$ & $0.57^{+0.05}_{-0.06}$ & $2.45^{+0.07}_{-0.07}$\\
                      & 9809C & S\'ersic & $41^{+3}_{-3}$ & $0.082^{+0.007}_{-0.006}$ & $0.49^{+0.04}_{-0.04}$ & $0.76^{+0.19}_{-0.07}$ & $0.68^{+0.05}_{-0.06}$ & $2.86^{+0.06}_{-0.07}$\\
\hline
\multirow{3}{*}{F150W} & 9809  & S\'ersic & $15^{+2}_{-1}$ & $0.12^{+0.03}_{-0.02}$ & $0.7^{+0.2}_{-0.1}$ & $0.9^{+1.2}_{-0.2}$ & $0.4^{+0.2}_{-0.2}$ & $1.2^{+0.5}_{-0.5}$\\
                      & 9809B & S\'ersic & $51^{+3}_{-2}$ & $0.068^{+0.004}_{-0.005}$ & $0.41^{+0.03}_{-0.03}$ & $0.72^{+0.11}_{-0.05}$ & $0.53^{+0.06}_{-0.07}$ & $2.52^{+0.09}_{-0.09}$\\
                      & 9809C & S\'ersic & $49^{+3}_{-3}$ & $0.071^{+0.006}_{-0.004}$ & $0.42^{+0.04}_{-0.02}$ & $0.80^{+0.23}_{-0.07}$ & $0.67^{+0.06}_{-0.07}$ & $2.86^{+0.06}_{-0.06}$\\
\hline
\multirow{3}{*}{F200W} & 9809  & S\'ersic & $56^{+4}_{-3}$ & $0.083^{+0.008}_{-0.007}$ & $0.50^{+0.05}_{-0.04}$ & $0.67^{+0.05}_{-0.01}$ & $0.24^{+0.10}_{-0.10}$ & $1.2^{+0.3}_{-0.3}$\\
                      & 9809B & S\'ersic & $43^{+3}_{-2}$ & $0.055^{+0.004}_{-0.005}$ & $0.33^{+0.03}_{-0.03}$ & $0.78^{+0.26}_{-0.10}$ & $0.54^{+0.08}_{-0.08}$ & $2.4^{+0.1}_{-0.1}$\\
                      & 9809C & S\'ersic & $55^{+3}_{-2}$ & $0.071^{+0.005}_{-0.004}$ & $0.42^{+0.03}_{-0.02}$ & $0.69^{+0.05}_{-0.02}$ & $0.62^{+0.05}_{-0.05}$ & $2.74^{+0.05}_{-0.06}$\\
\hline
\multirow{3}{*}{F277W} & 9809  & S\'ersic & $233^{+7}_{-5}$ & $0.102^{+0.003}_{-0.002}$ & $0.61^{+0.02}_{-0.01}$ & $1.19^{+0.05}_{-0.04}$ & $0.41^{+0.01}_{-0.01}$ & $1.72^{+0.02}_{-0.02}$\\
                      & 9809B & S\'ersic & $63^{+4}_{-3}$ & $0.2778^{+0.0007}_{-0.0008}$ & $1.667^{+0.004}_{-0.005}$ & $7.3^{+0.1}_{-0.2}$ & $0.032^{+0.001}_{-0.001}$ & $1.16^{+0.04}_{-0.05}$\\
                      & 9809C & S\'ersic & $88^{+4}_{-3}$ & $0.29913^{+0.00006}_{-0.00008}$ & $1.7946^{+0.0004}_{-0.0005}$ & $7.68^{+0.10}_{-0.16}$ & $0.00142^{+0.00012}_{-0.00010}$ & $1.3^{+1.3}_{-0.8}$\\
\hline
\multirow{3}{*}{F356W} & 9809  & S\'ersic & $541^{+7}_{-5}$ & $0.064^{+0.001}_{-0.001}$ & $0.384^{+0.008}_{-0.008}$ & $2.21^{+0.09}_{-0.08}$ & $0.29^{+0.02}_{-0.01}$ & $1.77^{+0.03}_{-0.02}$\\
                      & 9809B & S\'ersic & $36^{+2}_{-1}$ & $0.2816^{+0.0006}_{-0.0007}$ & $1.689^{+0.003}_{-0.004}$ & $6.3^{+0.3}_{-0.5}$ & $0.033^{+0.001}_{-0.001}$ & $0.8^{+0.2}_{-0.2}$\\
                      & 9809C & S\'ersic & $79^{+3}_{-3}$ & $0.29927^{+0.00004}_{-0.00006}$ & $1.7955^{+0.0002}_{-0.0004}$ & $7.6^{+0.1}_{-0.2}$ & $0.0019^{+0.0002}_{-0.0001}$ & $1.4^{+1.0}_{-0.8}$\\
\hline
\multirow{3}{*}{F410M} & 9809  & S\'ersic & $960^{+10}_{-10}$ & $0.036^{+0.001}_{-0.001}$ & $0.214^{+0.009}_{-0.009}$ & $3.3^{+0.3}_{-0.3}$ & $0.28^{+0.03}_{-0.03}$ & $1.69^{+0.06}_{-0.07}$\\
                      & 9809B & S\'ersic & $38^{+6}_{-4}$ & $0.24^{+0.02}_{-0.03}$ & $1.4^{+0.1}_{-0.2}$ & $0.70^{+0.14}_{-0.04}$ & $0.86^{+0.03}_{-0.09}$ & $1.03^{+0.07}_{-0.07}$\\
                      & 9809C & S\'ersic & $29^{+4}_{-4}$ & $0.06^{+0.03}_{-0.01}$ & $0.35^{+0.15}_{-0.08}$ & $3^{+3}_{-2}$ & $0.7^{+0.1}_{-0.3}$ & $2.6^{+0.3}_{-0.3}$\\
\hline
\multirow{3}{*}{F444W} & 9809  & S\'ersic & $1801^{+9}_{-7}$ & $0.0227^{+0.0006}_{-0.0007}$ & $0.136^{+0.004}_{-0.004}$ & $4.0^{+0.1}_{-0.1}$ & $0.26^{+0.03}_{-0.02}$ & $1.87^{+0.04}_{-0.04}$\\
                      & 9809B & S\'ersic & $33.0^{+1.2}_{-0.8}$ & $0.2789^{+0.0005}_{-0.0008}$ & $1.673^{+0.003}_{-0.005}$ & $5^{+1}_{-1}$ & $0.064^{+0.005}_{-0.003}$ & $1.9^{+0.5}_{-1.2}$\\
                      & 9809C & S\'ersic & $53^{+3}_{-3}$ & $0.2987^{+0.0001}_{-0.0001}$ & $1.7922^{+0.0007}_{-0.0009}$ & $7.3^{+0.4}_{-0.8}$ & $0.36^{+0.05}_{-0.07}$ & $2.1^{+0.7}_{-1.5}$\\
\hline
\end{tabularx}
\end{table*}

%% file: table_spec_fit.tex
\begin{table}
  \setlength{\tabcolsep}{2.5pt}
  \caption{Posterior probabilities for key parameters of the emission-lines model,
  based on simultaneous fits of the prism and G395M RUBIES spectroscopy. The five
  groups of rows are, from top to bottom, narrow-line parameters, outflow parameters,
  broad-line parameters, sodium-absorption parameters, and derived quantities.
  }\label{t.pars}
  \begin{tabular}{lcc}
\hline
Parameter                                        & Unit                                            & Posterior                \\
\hline
$z_\mathrm{n}$                                   & \hspace{-0.3cm}[---]                                           & $5.704^{+0.001}_{-0.001}$ \\
$\sigma_\mathrm{n}$                              & \hspace{-0.3cm}$[\mathrm{km\,s^{-1}}]$                         & $110^{+110}_{-70}$ \\
$A_V$                                            & \hspace{-0.3cm}$[\mathrm{mag}]$                                & $3.6^{+0.2}_{-0.2}$ \\
$F_\mathrm{n}(\Hbeta)$                  & \hspace{-0.3cm}$[10^{-18} \, \mathrm{erg\,s^{-1}\,cm^{-2}}]$   & $0.18^{+0.09}_{-0.08}$ \\
$F_\mathrm{n}(\OIIIL)$               & \hspace{-0.3cm}$[10^{-18} \, \mathrm{erg\,s^{-1}\,cm^{-2}}]$   & $0.12^{+0.14}_{-0.08}$ \\
$F_\mathrm{n}(\Halpha)$                 & \hspace{-0.3cm}$[10^{-18} \, \mathrm{erg\,s^{-1}\,cm^{-2}}]$   & $1.9^{+1.0}_{-0.9}$ \\
$F_\mathrm{n}(\SIIL[6716])/F_\mathrm{n}(\SIIL[6731])$ & \hspace{-0.3cm}[---]                          & $0.9^{+0.3}_{-0.3}$ \\
$F_\mathrm{n}(\OIIL[7330])/F_\mathrm{n}(\OIIL[7320])$ & \hspace{-0.3cm}[---]                          & $0.50^{+0.03}_{-0.02}$ \\
\hline
$F_\mathrm{out}(\OIIIL)$  & \hspace{-0.3cm}$[10^{-18} \, \mathrm{erg\,s^{-1}\,cm^{-2}}]$   & $4.4^{+0.3}_{-0.3}$ \\
$v_\mathrm{out}$                                 & \hspace{-0.3cm}$[\mathrm{km\,s^{-1}}]$                         & $-800^{+100}_{-100}$ \\
$\sigma_\mathrm{out}$                            & \hspace{-0.3cm}$[\mathrm{km\,s^{-1}}]$                         & $870^{+80}_{-70}$ \\
\hline
$v_\mathrm{BLR}$                                 & \hspace{-0.3cm}$[\mathrm{km\,s^{-1}}]$                         & $190^{+50}_{-70}$ \\
$FWHM_\mathrm{BLR,1}$                            & \hspace{-0.3cm}$[\mathrm{km\,s^{-1}}]$                         & $1200^{+200}_{-100}$ \\
$FWHM_\mathrm{BLR,2}$                            & \hspace{-0.3cm}$[\mathrm{km\,s^{-1}}]$                         & $3800^{+400}_{-300}$ \\
$F_\mathrm{BLR}(\Hbeta)$                & \hspace{-0.3cm}$[10^{-18} \, \mathrm{erg\,s^{-1}\,cm^{-2}}]$   & $1.6^{+0.2}_{-0.1}$ \\
$F_\mathrm{BLR}(\Halpha)$               & \hspace{-0.3cm}$[10^{-18} \, \mathrm{erg\,s^{-1}\,cm^{-2}}]$   & $66^{+1}_{-2}$ \\
$F_\mathrm{BLR,1}/F_\mathrm{BLR}$                & \hspace{-0.3cm}[---]                                           & $0.43^{+0.07}_{-0.05}$ \\
\hline
$v_\mathrm{abs}$                                 & \hspace{-0.3cm}$[\mathrm{km\,s^{-1}}]$                         & $-400^{+200}_{-100}$ \\
$\sigma_\mathrm{abs}$                            & \hspace{-0.3cm}$[\mathrm{km\,s^{-1}}]$                         & $300^{+100}_{-100}$ \\
$C_{f, \NaI}$                                  & \hspace{-0.3cm}[---]                                           & $0.5^{+0.2}_{-0.1}$ \\
$\tau_{\NaI}$                            & \hspace{-0.3cm}[---]                                           & $0.8^{+1.0}_{-0.4}$ \\
\hline
$F_\mathrm{BLR}(\mathrm{H\alpha})$               & \hspace{-0.3cm}$[10^{-18} \, \mathrm{erg\,s^{-1}\,cm^{-2}}]$   & $900^{+100}_{-100}$ \\
$FWHM_\mathrm{BLR}$                              & \hspace{-0.3cm}$[\mathrm{km\,s^{-1}}]$                         & $1500^{+200}_{-100}$ \\
$\log\,SFR(\mathrm{H\alpha})$                    & \hspace{-0.3cm}$[\mathrm{M_\odot\,yr^{-1}}]$                   & $1.3^{+0.2}_{-0.3}$ \\
$\log\,L_\mathrm{b}(\mathrm{H\alpha})$           & \hspace{-0.3cm}$[\mathrm{10^{42}\,erg\,s^{-1}}]$               & $2.54^{+0.05}_{-0.07}$ \\
$\log\,(M_\bullet)$                              & \hspace{-0.3cm}$[\mathrm{M_\odot}]$                            & $8.16^{+0.11}_{-0.09}$ \\
$\lambda_\mathrm{Edd}$                           & \hspace{-0.3cm}[---]                                           & $2.4^{+0.6}_{-0.5}$ \\
  \hline
    \end{tabular}
\end{table}

%% file: table_spectral_models.tex
\begin{table}
  \small
  \setlength{\tabcolsep}{2pt}
  \caption{Sensitivity of the emission-line \Av and SFR to the model assumptions. Unless otherwise specified, the models use the SMC-bar attenuation law and a double-Gaussian BLR, with a prior on $\Av<1.8/0.44$~mag derived from \cigale. We do not report the model uncertainties, since they are smaller than the model-to-model variations.}
  \label{t.modelgrid}
  \begin{tabular}{lrrr}
\hline
Model                                                & BIC      & $A_V$             & $\log\,SFR(\Halpha)$ \\
                                                     &          & [mag]             & [\Msun\,\peryr]      \\
\hline
$F_\mathrm{BLR}(\Halpha)/F_\mathrm{BLR}(\Hbeta)=10$  & $-589.7$ & $3.6$             & $1.3$ \\
$F_\mathrm{BLR}(\Halpha)/F_\mathrm{BLR}(\Hbeta)=5$   & $-587.4$ & $4.7$             & $1.7$ \\
$F_\mathrm{BLR}(\Halpha)/F_\mathrm{BLR}(\Hbeta)=3.1$ & $-565.6$ & $5.5$             & $2.3$ \\
No \cigale-based prior                               & $-578.5$ & $5.9$             & $2.2$ \\
Separate ISM/BLR $A_V$                               & $-566.8$ & $2.6/5.8^\dagger$ & $1.0$ \\
No $\mathcal{H}$ function                            & $-516.1$ & $5.4$             & $2.7$ \\
\FeII templates$^\ddag$                              & $-548.4$ & $5.3$             & $2.1$ \\
Single-Gaussian BLR                                  & $-389.8$ & $4.8$             & $3.4$ \\
Calzetti dust law                                    & $-471.4$ & $5.7$             & $2.0$ \\
\hline
  \end{tabular}
\raggedright{
$^\dagger$ For this model we report $A_{V,\mathrm{ISM}}$ and $A_{V,\mathrm{BLR}}$.
$^\ddag$ See Appendix~\ref{a.feii}.
}
\end{table}